\documentclass[12pt,preprint]{aastex}

\newcommand{\Mj} {M_{\rm J}}
\newcommand{\Msini}{M \sin{i}}
\newcommand{\n}{\phantom{0}}
\newcommand{\degree}{{}^{\circ}}

\shortauthors{Hajian et al.}
\shorttitle{Initial Results from the USNO Dispersed FTS}

\begin{document}

\title{Initial Results from the \\ 
USNO Dispersed Fourier Transform Spectrograph}

\author{Arsen R.\ Hajian\altaffilmark{1},
	Bradford B. Behr\altaffilmark{1,2},
	Andrew T.\ Cenko\altaffilmark{1},
	Robert P.\ Olling\altaffilmark{1,3},
	David Mozurkewich\altaffilmark{4},
	J.\ Thomas Armstrong\altaffilmark{2}, 
	Brian Pohl\altaffilmark{1, 5},
	Sevan Petrossian\altaffilmark{1,6},
	Kevin H.\ Knuth\altaffilmark{7},
	Robert B.\ Hindsley\altaffilmark{2},
	Marc Murison\altaffilmark{1}, 
	Michael Efroimsky\altaffilmark{1}, 
	Ronald Dantowitz\altaffilmark{8}, 
	Marek Kozubal\altaffilmark{8},
	Douglas G.\ Currie\altaffilmark{9},
	Tyler E.\ Nordgren\altaffilmark{10,11},
	Christopher Tycner\altaffilmark{6,10},
	Robert S.\ McMillan\altaffilmark{12}\\
email: {\it hajian@usno.navy.mil, bbb@usno.navy.mil,
atc@usno.navy.mil, olling@astro.umd.edu, dave@mozurkewich.com,
tom.armstrong@nrl.navy.mil, bpohl@physics.unc.edu, spetrossian@excelgov.org,
kknuth@albany.edu, hindsley@nrl.navy.mil, murison@usno.navy.mil,
me@usno.navy.mil, dantowitz@dexter.org, marek@portents.com,
currie@umd.edu, tyler\_nordgren@redlands.edu, tycner@nofs.navy.mil,
bob@lpl.arizona.edu}}

\altaffiltext{1}{\small US Naval Observatory, 3450 Massachusetts Av,
NW, Washington, DC 20392-5420.}
	
\altaffiltext{2}{\small Remote Sensing Division, Naval Research
Laboratory, Code 7215, Washington, DC, 20375. }
	
\altaffiltext{3}{\small Department of Astronomy, University of Maryland,
College Park, MD 20742-2421.}
	
\altaffiltext{4}{\small Seabrook Engineering, 9310 Dubarry Avenue, Seabrook, MD 20706.}

\altaffiltext{5}{\small Department of Physics and Astronomy, University of
North Carolina at Chapel Hill, Phillips Hall, CB \#3255, Chapel Hill,
NC 27599-3255.}

\altaffiltext{6}{\small NVI, 7257D Hanover Pkwy., Greenbelt, MD 20770.}

\altaffiltext{7}{\small Intelligent Systems Division, NASA Ames
Research Center, Moffett Field CA 94035 (Currently at: Department of Physics,
University of Albany (SUNY), Albany NY 12222).}

\altaffiltext{8}{\small Clay Center Observatory, Dexter and Southfield
Schools, 20 Newton St., Brookline, MA 02445.}

\altaffiltext{9}{\small Department of Physics, University of Maryland,
College Park, MD 20742.}

\altaffiltext{10}{\small US Naval Observatory, Flagstaff Station,
10391 W.\ Naval Observatory Rd., Flagstaff, AZ, 86001.}

\altaffiltext{11}{\small Department of Physics,
University of Redlands, 1200 East Colton Avenue, P.O. Box 3080,
Redlands, CA 92373.}

\altaffiltext{12}{\small Lunar and Planetary Laboratory, University of
Arizona, Tucson, Arizona 85721.}


\begin{abstract}
We have designed and constructed a ``dispersed Fourier Transform
Spectrometer'' (dFTS), consisting of a conventional FTS followed by a
grating spectrometer. By combining these two devices, we negate a
substantial fraction of the sensitivity disadvantage of a conventional
FTS for high resolution, broadband, optical spectroscopy, while
preserving many of the advantages inherent to interferometric
spectrometers. In addition, we have implemented a simple and
inexpensive laser metrology system, which enables very precise
calibration of the interferometer wavelength scale. The fusion of
interferometric and dispersive technologies with a laser metrology
system yields an instrument well-suited to stellar spectroscopy,
velocimetry, and extrasolar planet detection, which is competitive with
existing high-resolution, high accuracy stellar spectrometers. In this
paper, we describe the design of our prototype dFTS, explain the
algorithm we use to efficiently reconstruct a broadband spectrum from
a sequence of narrowband interferograms, and present initial observations
and resulting velocimetry of stellar targets.
\end{abstract}



\section{Introduction}

The past two decades have seen a tremendous improvement in the
capabilities of astronomical spectrometers. Velocity precisions of
1~km/s were rarely achieved prior to 1980, while the current
generation of high-precision spectrometers boast precisions of a few
m/s or less. Such instruments have been able to find planetary
companions with $0.1 \Mj < \Msini < 15 \Mj$ in over 150
stellar systems (where $i$ is the inclination angle of the orbit of
the companion, and $\Mj$ is the mass of Jupiter), by detecting
periodic variation in the stellar radial velocity (RV).

Summaries of the advantages and limitations of the spectroscopic
instrumentation and data reduction procedures are discussed elsewhere
(Butler {\it et al.}\ 1996; Marcy \& Butler 1998; Baranne {\it et al.}\
1996). The majority of the planet detections to date have been made using
cross-dispersed echelle spectrometers equipped with molecular iodine
absorption cells. The gas absorption lines provide a wavelength reference
scale, superposed on each observed stellar spectrum, which is sufficiently
stable to give long-term radial velocity accuracies as small as 1~m/s. More
recently, thorium-argon emission line calibration has gained popularity,
and also achieves 1~m/s accuracy (Lovis {\it et al.}\ 2006). These
precision wavelength calibration techniques have been refined by a number of
different research groups, and are now in widespread use.  However, if
planetary masses significantly smaller than $\Mj$ are to be inferred
spectroscopically, or if other spectroscopic studies requiring instrumental
precisions better than $\approx 1$~m/s are needed, it is likely that a very
different type of instrument is required.

Echelle spectrographs are not the only option for high-precision,
high-resolution spectroscopy. For many observations, a Fourier
Transform Spectrograph (FTS) provides superior resolving power and
wavelength accuracy (Brault 1985). The concept of the FTS has been
around for over a century; the theoretical basis was laid at the end
of the $19^{\rm th}$ century (Michelson 1891, 1892), but FTSs did not
achieve widespread use until $\approx$75~years later (Ridgway \&
Brault 1984 and references therein), after requisite technological
advances in optics, precise motion control, and laser metrology were
made. FTS devices are now common for laboratory spectroscopy,
atmospheric sensing, and numerous other applications. However,
astronomers have yet to embrace interferometric spectrometers for the
purpose of obtaining precise velocities of stars in the optical
regime. (For spectroscopy at radio wavelengths, by contrast, the FTS
is the standard device of choice.)

The unpopularity of the FTS for optical spectroscopy is well-founded:
interferometric devices are generally more complex, mechanically
demanding, and cumbersome than their grating spectrometer
counterparts. But most importantly, the effective throughput of a
conventional FTS observing a broadband source in the
photon-noise-limited regime is inferior compared to a conventional
spectrometer with an array detector. To achieve good fringes over a
range of delays, the bandwidth of an FTS is typically restricted to a
narrow slice of the optical spectrum, and for broadband use, a
scanning FTS, which collects one interferogram data point at each
delay position, has an efficiency equivalent to a single-pixel
scanning spectrometer or monochromator. This drawback has prevented
the FTS from being widely used in astronomy at optical wavelengths,
where sensitivity is of paramount importance. For precise stellar
radial velocity measurements, spectral resolutions of $\approx$50,000
are required in order to resolve stellar absorption lines. A
broadband, conventional, photon-limited FTS operating at this
resolution will convert detected photons into spectral signal-to-noise
ratio with an efficiency 50,000 times smaller than a dispersive
spectrograph.  For all the advantages of FTS devices, their poor
efficiency renders them essentially useless for precise stellar
velocimetry. As a result of this limited sensitivity, FTSs are
commonly used only in situations where sensitivity can be sacrificed
for precision, such as laboratory spectroscopy (Kerber {\it et al.}\
2006; Aldenius, Johansson, \& Murphy 2006; Ying {\it et al.}\ 2005),
solar observations (Fawzy, Youssef, \& Engvold 1998), or where very
high spectral resolution or accurate wavelength calibration are
required, such as in measurements of planetary atmospheres (Cooper
{\it et al.}\ 2001; Krasnopolsky, Maillard, \& Owen 2004).

The key to surmounting this limitation of the traditional FTS is to
divide the broad spectral bandpass into many narrow-bandpass channels,
by placing a dispersive grating at the output of the interferometer,
and then focusing the resulting medium-resolution spectrum onto an
array detector. Each pixel on the detector sees only a tiny range of
wavelengths, so the interferometric fringes remain visible with a high
signal-to-noise ratio over a much wider range of optical path
difference, and the interferograms can be sampled much more coarsely
without sacrificing information. In essence, by adding a
post-disperser, we have created a few thousand separate narrow-band
FTSs, all operating in parallel. We have named the resulting device
the ``dispersed Fourier Transform Spectrograph,'' or {\it dFTS}.

Combinations of FTS and dispersive technologies have been considered
or developed by other instrumentation projects. Jennings {\it et al.}\
(1986) briefly describe using a grating monochromator at the output of
the Kitt Peak 4-meter facility FTS to select specific narrowband
output channels (albeit without multiplexing). Mosser, Maillard, \&
Bouchy (2003) discuss the advantages of using a low-resolution
dispersive element to collect multiple parallel interferograms in
simulations of an FTS-based asteroseismology spectrograph.  In a
similar vein, the ``Externally Dispersed Interferometer'' (EDI)
concept described by Erskine (2003) and Erskine, Edelstein, \&
Feuerstein (2003) uses a Michelson interferometer to induce spectral
fringes on a high-resolution optical spectrum, providing wavelength
calibration and boosting the spectral resolution by a factor of 2 to
3. An EDI-based device recently discovered a new planet (Ge {\it et
al.} 2006), demonstrating the potential of spectral interferometry for
stellar velocimetry. Both the Mosser {\it et al.} and Erskine {\it et
al.} concepts operate at a fixed non-zero delay position, or scanning
over a small range of closely-spaced delays, whereas the dFTS coarsely
samples the interferogram over a wide range of delay positions, so
that a complete high-resolution broadband spectrum can be
reconstructed.

Another advantage of the dFTS design lies in its built-in laser
metrology system. In order to accurately reconstruct the input
spectrum from the measured interferograms, we must precisely measure
the optical path difference (OPD) between the two arms of the
interferometer while fringe data are being acquired. We send a
collimated polarization-modulated beam from a frequency-stabilized
laser through the same interferometer path as the starlight beam, from
splitting to recombination, and then extract the laser signal to
measure OPD changes with an accuracy of $\approx 0.1$~nm. By
continuously monitoring the path length difference during data
acquisition via the metrology system, we can unambiguously assign a
path length difference, or delay, to each fringe measurement of the
dFTS, preserving the wavelength scale of the resulting spectrum with
high precision (we use the term ``lag'' interchangeably with
``delay'').  This wavelength solution extends across the entire
optical bandpass, unlike the iodine calibration technique, which loses
effectiveness for $\lambda < 510$~nm.

Yet another virtue of the interferometric approach is that the
instrument can be smaller and cheaper than an echelle with equivalent
resolving power.  This is particularly important for telescopes with
large apertures. Since the interferometer section of the dFTS provides
all of the high-resolution spectral capability, the demands on the
dispersing element are greatly relaxed, and the collimated beam
diameter within the dispersive spectrograph section can be an order of
magnitude smaller than otherwise. Smaller optics will be considerably
less expensive, and the overall instrument size can be as large as a
desk instead of occupying an entire room.

With these instrumental modifications, the FTS has the potential of
surpassing the $\approx 1$~m/s accuracy achieved by absorption-cell
echelle spectrographs, and doing so with smaller and cheaper hardware.
At this level of RV precision, apparent Doppler velocity oscillations
can be induced by the convective and turbulent motions of the star's
surface, even for relatively old, inactive stellar types. These
sources of ``astrophysical noise'' pose a significant challenge to
detection of low-mass planets.

In this paper, we describe the design, construction, and testing of a
prototype dFTS. In \S 2, we compare the optical configurations of a
conventional FTS and a dFTS, and describe the hardware implementation
of our dFTS instrument. The data acquisition and processing systems
are detailed in \S 3. Initial results on calibration light sources,
which test the systematic error limits of the device, are shown in \S
4. Radial velocity measurements on stellar targets, including
spectroscopic binaries and exoplanet systems, are presented in \S 5,
along with a discussion of the precision limits of the instrument. In
\S 6, we summarize the status and initial results of the current dFTS,
and explore future prospects for this technology. Appendices~A and B
then describe the theory of spectral multiplexing and our FROID
(Fourier Reconstruction of Optical Interferometer Data) algorithm in
detail.


\section{Interferometer Design}

\subsection{Review of Conventional FTS}

\begin{figure}
\epsscale{0.70}
\plotone{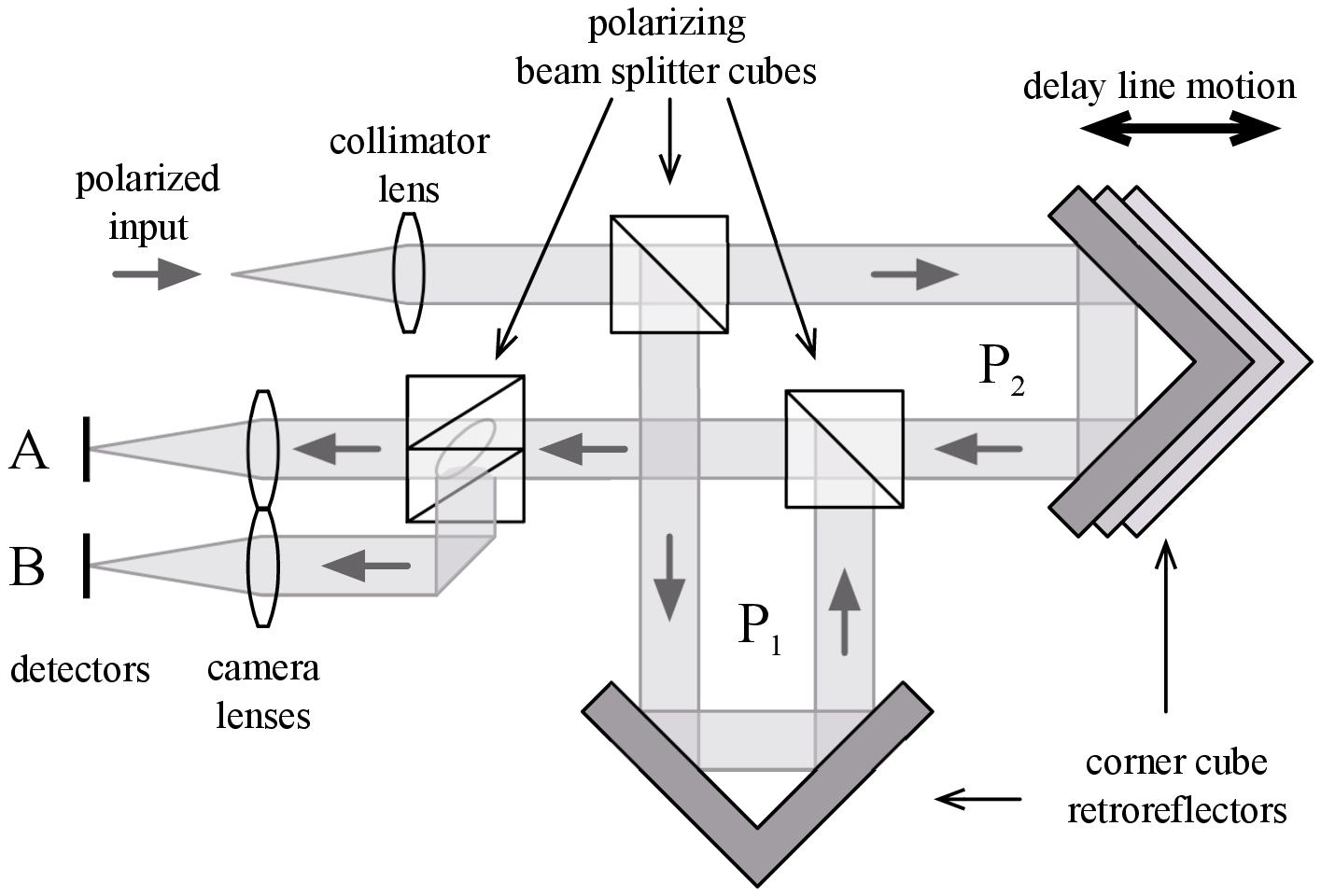}
\caption{A schematic layout of a conventional FTS, using polarizing
optics.  Polarizing beamsplitters BS1 and BS4 are mounted at
$45^{\circ}$ to the plane of the figure, so that one diagonal linear
polarization is transmitted while the orthogonal diagonal polarization
is reflected.}
\label{conventional-fts}
\end{figure}

In Figure~\ref{conventional-fts}, we show a cartoon layout of a
conventional white-light FTS using polarizing optics. The progress of
the light beam through this apparatus can be outlined in five stages:
(1) The incoming collimated beam of light is divided into two beams by
a polarizing beam splitter cube (PBSC). (2) The beams follow separate
paths, $P_1$ and $P_2$. The length of $P_2$, the ``delay line,'' can
be precisely adjusted by translating the moving retroreflector. (3)
The beams are recombined by another PBSC. (4) Using a third PBSC, the
recombined beam is split into two orthogonal diagonal polarizations to
induce fringing.  (5) The intensity of each of the recombined beams
(A, B) is separately measured by a detector for a sequence of
different delay line positions.  As is the case for most
interferometers, these two outputs are complementary (e.g. their
fringes will be 180$^{\rm o}$ out of phase from each other to conserve
energy).  The following discussion refers only to output A.

The wavelengths in the incoming light beam cover a range from
$\lambda_{\rm min}$ to $\lambda_{\rm max}$, i.e., centered on
$\lambda_0$ and covering a range $\Delta\lambda = \lambda_{\rm max} -
\lambda_{\rm min}$. The most important length parameter in the FTS is
the {\it delay}, $x$, which is equal to the optical path length
difference between paths $P_1$ and $P_2$. At any given wavelength
$\lambda$, complete constructive interference between light from the
two paths occurs when $x/\lambda$ is an integer, and complete
destructive interference occurs when $x/\lambda$ is an integer plus
$1/2$. When the paths $P_1$ and $P_2$ are precisely equal to within a
small fraction of $\lambda_0$ (i.e., $x/\lambda = 0$ at all
wavelengths), the intensity $I$ of the output beam is at its maximum,
$I_{\rm max}$, since the light waves at all wavelengths constructively
interfere.  This position is known as the {\it central fringe}.  As we
move the delay line and $x$ increases, the interference fringes weaken
and $I$ decreases. As $x$ continues to increase, $I$ reaches a minimum
at $x/\lambda_0 = 1/2$ and then rises again to a new (but weaker)
maximum at $x/\lambda_0 = 1$. This weakening oscillation of $I$
continues as $x$ increases. When $x/\lambda_0$ is many times larger
than $\lambda_0/{\Delta\lambda}$, some wavelengths interfere
constructively and some destructively, so $I$ is close to the mean
light level (i.e. 0.5 $I_{\rm max}$).  Thus, if the observed spectral
region $\Delta\lambda$ is wide, there is only a small range of delay
near the central fringe with large deviations from the mean level for
both output beams.

\begin{figure}
\epsscale{0.80}
\plotone{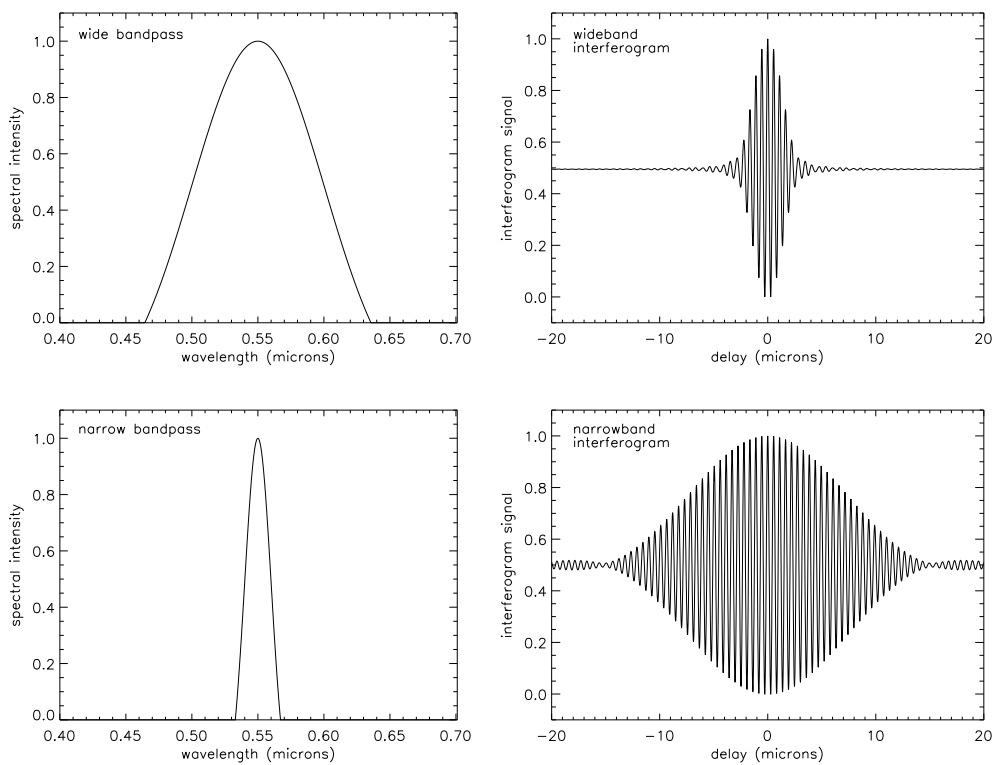}
\caption{Simulated interferograms, to illustrate the relationship
between spectral bandpass and fringe packet size. The wavelength of
the high-frequency oscillations is the central wavelength of the
bandpass, $\lambda_{\rm o}$. The number of fringes in the central
fringe packet is approximately equal to $\lambda_{\rm
o}/\Delta\lambda$ where $\Delta\lambda$ is the FWHM of the spectral
bandpass.}
\label{fringe-packets}
\end{figure}

The resulting data set of intensity measurements at many values of $x$
is known as an {\it interferogram}. The region of $x$ over which there
are large deviations from the mean level is termed the {\it fringe
packet}. Illustrative examples of interferograms are shown in
Figure~\ref{fringe-packets}. In the limit of infinite bandwidth, the
Nyquist Theorem requires sampling the interferogram in steps of
$\lambda_{\rm min}/2$ in order to avoid losing spectral information.
In practice, the interferogram is often sampled somewhat more finely
than this. The resolution of the spectrum is determined by the maximum
value of $x/\lambda$. We can see this if we imagine a spectrum
consisting of a single narrow emission feature.  Its interferogram
will have a large range of $x$ over which $I$ oscillates. In order to
see just how narrow the spectral feature is, we must continue to
increase $x$ until the oscillations in $I$ diminish.

In principle, an FTS offers three advantages over a dispersing
spectrometer. (1) The spectral resolution can be changed simply by
changing the maximum value of $P_1-P_2$, i.e. the delay line scan
range.  (2) The wavelength scale in the resulting spectrum is computed
directly from the delay line measurements, and is insensitive to such
effects as scattered light and flexure of the instrument. (3) The line
spread function (instrumental broadening function) of the resulting
spectrum can be derived, to a high degree of precision, directly from
the delay sampling function.

However, as mentioned previously, traditional FTSs suffer from low
sensitivity, because much of the delay scan range produces a small
signal, and because a large number of measurements must be done
sequentially to produce a well-sampled interferogram.  Also, fringes
can be difficult to distinguish from temporal flux variations in the
source, so some flux normalization procedure must be implemented when
observing stars through the turbulent atmosphere.

\subsection{Dispersing the White Light Fringe}

Our strategy for mitigating these shortcomings is to use a grating
spectrometer to disperse the recombined white light beam emerging from
a conventional FTS. This technique converts a single, broadband FTS
into numerous narrowband interferometers, all functioning in
parallel. Since the width of the fringe packet is inversely
proportional to the spectral bandwidth, dispersing the white light
beam into narrowband channels serves to broaden the fringe packet for
each channel by a factor of $10^3$ or more.

There are two significant gains that are realized by the dispersed
interferometer. The first advantage results directly from the Nyquist
Theorem, which states that in order to avoid aliasing, the fringes
must be sampled at intervals, $\delta x$, that are at most:
	\begin{equation}
	\delta x < {1\over 2 \Delta s},
	\end{equation}
where $\Delta s$ is the bandwidth in wavenumber.  Since the
postdispersion narrows the bandwidth for a given spectral channel, the
fringes can be sampled at wider intervals. For a fixed total delay
range (i.e. spectral resolution), fewer samples are needed to
reconstruct an unaliased narrowband signal than an unaliased broadband
signal. 

Second, since the fringe packet for a narrowband channel is wider than
that of the broadband white light (recall
Figure~\ref{fringe-packets}), a larger fraction of the delay range is
spanned by high-contrast fringe signal.  As a result, the
signal-to-noise ratio in the resulting spectrum is increased.  We
prove this from basic principles in Appendix~A. More detailed
discussions, derivations, and specific examples of the advantages
resulting from dispersing the white light fringe can be found in
Appendix~B, where we discuss our spectral reconstruction algorithm in
detail.

\subsection{Instrument overview}

\begin{figure}
\epsscale{0.90}
\plotone{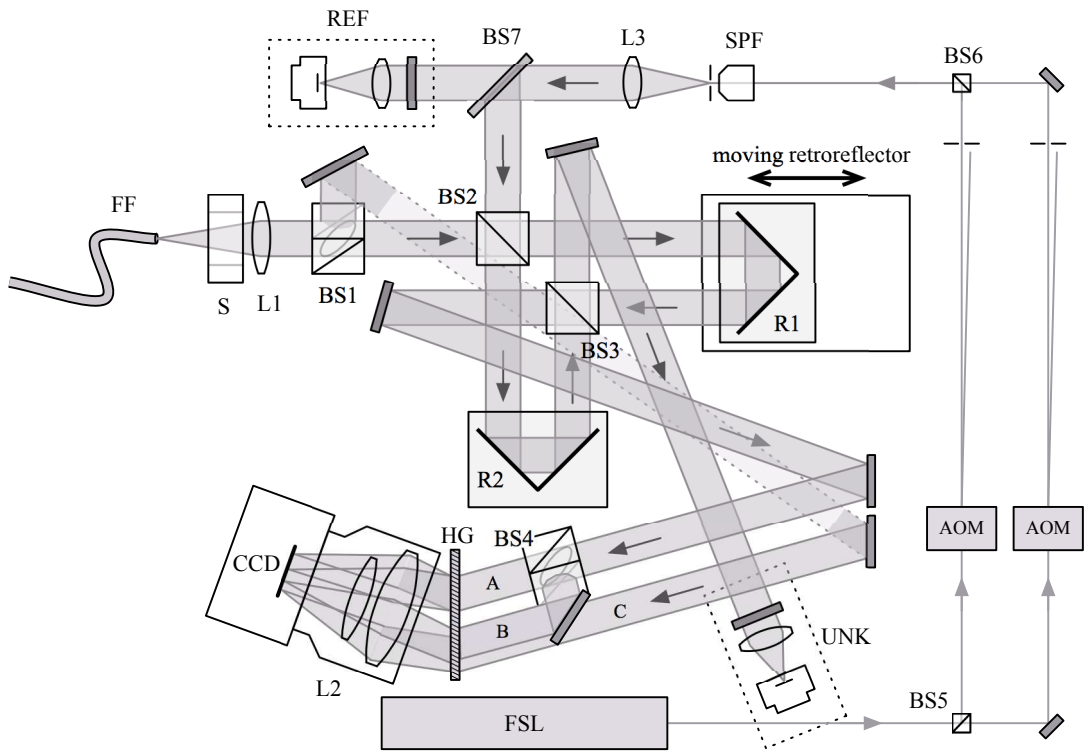}
\caption{A schematic drawing of the current instrumental configuration for the dFTS prototype.}
\label{dfts-layout}
\end{figure}

Our adopted configuration for the dispersed FTS prototype is shown in
Figure~\ref{dfts-layout}. Light is guided from the telescope to the
interferometer through a multimode optical fiber feed {\bf FF}. (The
implications of this mode of transport on the measured interferometric
correlation is discussed in \S 2.5.) The light passes through a
mechanical shutter {\bf S} (a Uniblitz VS25, with a nominal opening
time of 6 ms) and is then collimated with an achromatic doublet lens
{\bf L1} into a beam with a diameter of $\approx 23$~mm.

It is assumed that the electric field at any position in the light
beam has a random polarization vector.  Since the resulting
performance of an FTS is optimal when the input beam is split in two
equal portions, and since we intend to use a polarizing beamsplitter
(which reflects vertically polarized light and transmits horizontally
polarized light) for the interferometer, we must ensure that all of
the light reaching the interferometer is linearly polarized at
45$^{\rm o}$ to the plane of the table.  To achieve this polarization,
the light passes through {\bf BS1}, a PBSC oriented at $45\degree$
relative to the optical table. The light transmitted through this cube
is thus polarized at $45\degree$ relative to the axes of the
interferometer beamsplitters, while the light reflected from BS1 is
routed by two relay mirrors to the dispersive back-end of the
instrument, where it serves as an unfringed ``photometric'' signal for
flux normalization (the so-called C beam).

The beam then enters the interferometer proper, splitting into its
vertically and horizontally polarized components, the V beam and the H
beam, at PBSC {\bf BS2}. Each sub-beam travels down one arm of the
interferometer, enters a retroreflecting corner cube ({\bf R1} or {\bf R2})
consisting of three mirrors, and emerges parallel to the incoming sub-beam,
but displaced laterally by approximately 4~cm.  The H and V sub-beams meet
at {\bf BS3}, where they recombine. Since this ``beam-combiner'' cube is
also polarizing --- reflecting vertically-polarized light, and transmitting
horizontally-polarized light --- nearly all of the starlight emerges from
one face of BS3 (traveling leftwards in the figure).

At this point, the recombined beam contains two sub-beams, H and V.
In order to see interference patterns, we must send the beam through
yet another polarizing beamsplitter, {\bf BS4}, oriented at
$45\degree$ to the optical table plane (like {\rm BS1}). Half of each
of the H and V beams are transmitted by BS4 and emerge with a
$+45\degree$ diagonal linear polarization. We call this beam A. The
other half of the H and V beams are reflected by BS4 and have a
$-45\degree$ diagonal linear polarization. They are reflected by a
mirror into beam B, parallel to beam A. Since the photons in each of
beams A and B have the same polarization, they can interfere, and
broadband fringes would appear if we placed detectors within the beams
at this point.

\begin{figure}
\epsscale{0.80}
\plotone{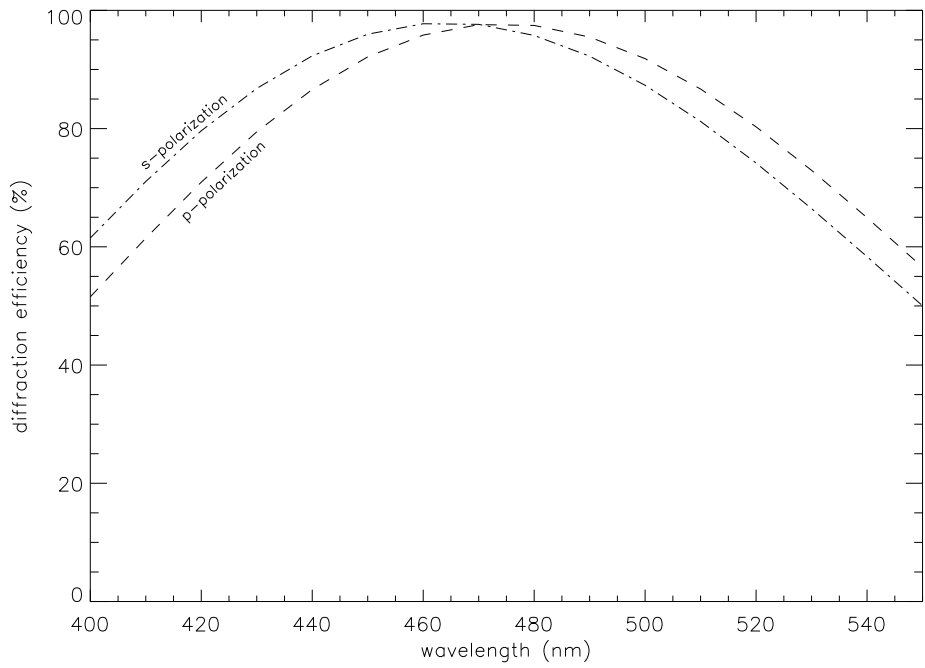}
\caption{The diffraction efficiency curves for the holographic transmission grating,
	as reported by the manufacturer.}
\label{grating-curve}
\end{figure}

Instead, we send both the A and B beams (plus the photometric C beam,
which did not pass through the interferometer section) into the
dispersive ``backend'' of the dFTS system. We are interested in
dispersing the white light to form spectra covering a wavelength range
of 460--560~nm, as the density of absorption lines in this region is
high for late-type stars, and thus is rich in radial velocity
information. Our choice for the dispersing element is a holographic
transmission grating {\bf HG}, manufactured by Kaiser Optical
Systems. We obtained a 10 cm $\times$ 10 cm grating with 1800
lines/mm, blazed to first order at $\lambda_0$ = 470 nm at an
angle of $25\degree$. The peak efficiency of these gratings is very
high, as illustrated by the transmission curves in
Figure~\ref{grating-curve}.

The dispersed beams are focused onto the detector by an f/2 Nikon
camera lens {\bf L2} (see Figure 3) with focal length 135~mm. The A,
B, and C beams enter the lens aperture tilted slightly relative to
each other so that the three spectral ``tracks'' do not overlap on the
detector. Our detector is an Andor DU-440 CCD ($2048 \times 512$
pixels, $13.5 \mu$m per pixel), which yields an average dispersion of
0.05 nm/pixel. The $50\mu$m input fiber diameter subtends several
pixels on the CCD, and from our calibration data, we derive a FWHM
bandpass per channel of 0.30 nm, so the spectral resolving power of
the grating backend is thus $R \approx 1700$. To prevent stray laser
reflections from entering the CCD, a shortpass filter with a 50\%
cutoff wavelength of 600~nm is placed in front of the CCD window.

\subsection{Metrology System}

In order to achieve the desired wavelength accuracy in the final
stellar spectra, we need to precisely measure the optical path
difference in the interferometer at each delay position.  We use a
laser metrology system to accomplish this.  The metrology beam follows
the same optical path through the interferometer as the starlight
does. We then employ a phase-locked loop (PLL) to track the metrology
fringes in real time. Instead of implementing the PLL in hardware, we
do most of the phase tracking in software, resulting in a simple,
inexpensive design that requires only off-the-shelf components. The
lack of custom signal acquisition boards (as is necessary in the case
of a hardware PLL) results in cost and time savings.

The metrology system begins with a frequency-stabilized helium-neon
(HeNe) laser, {\bf FSL} in Figure~\ref{dfts-layout}, operating at a
wavelength of 632.8~nm. We selected the Melles Griot 05-STP-901 laser,
which offers good wavelength stability and output power for a modest price.
A collimated beam with a diameter of $\approx 3$~mm emerges from the laser.
The laser tube is rotated so that the plane of polarization is oriented at
$45\degree$ relative to the plane of the optical table. The laser beam is
split by a polarizing beamsplitter cube {\bf BS5}, and each sub-beam passes
through a separate acousto-optical modulator (AOM), one of them driven at
40~MHz, the other at {40~MHz + 11~kHz}. (The same controller unit drives
both AOMs, to ensure that the 11 kHz frequency difference remains
constant.) The AOMs act like a ``moving grating'', with an unshifted
zero-order beam, flanked by first order beams with frequency shifts of $\pm
40$~MHz (for one AOM) and $\pm 40.011$~MHz (for the other AOM). We rotate
each AOM (in essence, adjusting the angle of incidence of the ``grating'')
to put as much power into the desired first-order beam as possible, and
block the other beams. Each AOM thus produces one beam; the two beams have
orthogonal linear polarizations and a frequency difference of 11 kHz.

These two laser beams are recombined at a second polarizing
beamsplitter cube {\bf BS6}, spatially filtered with a $20\times$
microscope objective and a $10\mu$ pinhole ({\bf SPF}), collimated
with lens {\bf L3}, and sent through an iris with a diameter of
$\approx 23$~mm, which is the same diameter as the starlight beam. The
resulting metrology beam contains horizontally-polarized light at one
frequency and vertically-polarized light at a slightly different
frequency. A non-polarizing beamsplitter {\bf BS7} is used to divert
one half of this beam to the ``reference detector cluster'' ({\bf
REF}), consisting of a sheet polarizer (oriented at $45\degree$),
followed by a lens which focuses the beam onto a Thorlabs PDA-55 PIN
photodiode detector. The polarizer mixes the H and V metrology beams,
producing an intensity modulation at 11 kHz, which is detected by the
photodiode and then transported via coaxial cable to one of the inputs
of a digital lock-in amplifier, the Stanford Research SR830.

The remaining metrology signal is injected into the interferometer,
entering BS2 orthogonal to the incoming white light beam. At BS2, the
two polarizations of the metrology beam are separated: one travels
through one arm, the other through the other arm. By adjusting the
input angle and position at BS2, the metrology beams are made to be
completely coincident in position and direction with the white light
beams, so that they pass through the same airmass and reflect off of
the same mirror surfaces within the interferometer section of the
instrument. This alignment, and the resulting full-aperture
metrology data, is crucial for accurately measuring the optical path
difference for the starlight beam.  Full aperture metrology
eliminates a large number of potential instrumental systematic errors,
which would otherwise plague the stellar spectra obtained with this
device.

At BS3, the metrology beams are recombined, and they exit the
interferometer orthogonal to the white light exit beam. The combined
beam is routed to the ``unknown detector cluster'' ({\bf UNK}), which
is identical to the reference detector cluster: a sheet polarizer at
$45\degree$, which causes the horizontal and vertical polarizations to
mix together, and a lens to focus the beam onto a second PIN detector,
where we detect a similar 11 kHz modulation. This UNK signal is sent
to the second input of the lock-in amplifier, where it is
phase-referenced to the REF reference signal (see \S 3.1).

\subsection{Fiber Input}

In order to transport photons from the telescope focal plane to the
spectrometer, we utilize a Ceramoptec Optran UV-50/125 multimode fiber
cable, 20 meters in length, with an armored jacketing to prevent
excessive bending or damage. This fiber has a core diameter of $50\mu$m
and a rated numerical aperture (NA) of 0.22. (Previous iterations of
the dFTS hardware utilized different fibers with lower NA, chosen in
order to control focal-ratio degradation, but the throughput of these
``slow fibers'' was correspondingly poor: 33\%\ to 50\%\ in some cases,
compared to $> 90$\% for the current fiber.)  Light from the telescope
typically enters the fiber at f/6.5, and it emerges at approximately
f/4, so that nearly all of it makes it through the initial collimating
lens and iris.

Keeping the star image centered on the fiber input aperture requires
feedback to the telescope steering system. We have constructed a
customized ``guider box,'' which bolts on to the Cassegrain port of
the telescope. The converging f/10 beam from the telescope secondary
is sped up to f/6.5 by a focal reducer lens, then passes through a 8\%
reflective pellicle. The beam comes to focus at the front face of the
fiber ferrule, where most of the starlight ``disappears'' down the
fiber aperture, which subtends 2.5~arcsec on the sky. A small fraction
of the light (the extreme wings of the seeing profile, which hits the
polished metal ferrule face outside the fiber cladding diameter, as
well as the $\approx 4$\% Raleigh reflection from the fiber core
region) bounces back up to the pellicle, and is reflected to a small
achromatic lens, which reimages the image from the fiber end onto an
Astrovid StellaCam~II video camera. We manually steer the telescope to
place the star on the fiber center, and then enable custom guiding
software, which adjusts the telescope pointing to keep the star image
centered on the fiber core based on feedback from the Astrovid camera.

Using multimode fiber results in good light gathering ability at the
telescope focal plane, and a scrambled wavefront emerging from the
waveguide. The scrambling of the wavefront would be fatal if we were
correlating light from spatially separated apertures (spatial
interferometry), but we are autocorrelating the light from a
single aperture (temporal interferometry). 

\subsection{Alignment system}

For maximum fringe contrast and metrology accuracy, the collimated
beams within the interferometer must be coincident in pupil position
to $<1$~mm, and parallel to within a few arcseconds. In order to
achieve these specifications, many of the key optical elements within
the instrument are under remote tip-tilt or X-Y translation control,
using New Focus ``Picomotor'' actuators. To align pupil positions, we
insert white cards into the collimated beams using electric-motor
actuators, and then view the pupil images using modified webcams. For
evaluating angular differences between beams, we use a Picomotor to
rotate a pick-off mirror into a collimated beam, rerouting it through
a long-focal-length lens and directly onto another webcam
detector. Using beam blockers to blink between two different
overlapping beams (e.g. metrology vs. starlight, or
interferometer arm~1 vs. arm~2), we can then evaluate any positional
or angular misalignments, and correct them via Picomotor. A Java-based
user interface controls all the alignment motors and cameras, and
walks the user through the proper alignment sequence. Using this
system, we achieve positional accuracy of $\sim 0.1$~mm and angular
accuracy of $\sim 4$~arcsec.

\subsection{Temperature stabilization}

The temperature of the dFTS optical elements, optomechanical mounts, and
breadboard must remain constant, to minimize nightly realignment.
This requirement is a particular challenge
at the prototype's current location, in a room which is vented to the
outside air to reduce dome seeing during observations. The entire dFTS
instrument is enclosed in a sturdy wooden crate, with 4-inch thick
Celotex insulation panels on all six faces. Air is circulated through
the enclosure in a closed cycle, exiting the box via insulated
flexible pipe and passing through a 400~BTU/hr thermoelectric air
conditioner and a 150~W heater, which are regulated by an Omega CNi8
temperature controller. Thermocouple probes are located throughout the
interior of the instrument box, to provide feedback signal to the
Omega controller, and allow us to monitor the instrument
temperature. This temperature control system has proven sufficient to
maintain internal air temperatures to $\pm 0.5\degree$F, even when the
room temperature drops to $25\degree$F (during winter observing) or
rises to $85\degree$F (during the summertime). We also monitor the
atmospheric pressure and relative humidity within the box, and use these
data to correct for atmospheric dispersion effects within the interferometer.


\section{Data Acquisition}

In order to minimize the complexity of the control software, we wanted
to avoid a realtime operating system for the dFTS computer
systems. Fortunately, the adopted hardware configuration and observing
logistics permitted us to adequately control the hardware with two
standard 450 MHz PC computers running the Microsoft Windows 2000
operating system. We name these the ``metrology'' and ``fringe''
computers, and describe their functions and interactions below.

\subsection{Metrology Data}

The lock-in amplifier (LIA) takes the analog signals from the REF and
UNK metrology detectors as its inputs. Within the LIA, signals are
passed through an analog-to-digital converter and mathematically
analyzed by a digital signal processor. The phase difference between
the two signals at the carrier frequency is isolated by the LIA,
suppressing noise at other frequencies and resulting in a high
signal-to-noise ratio detection. The outputs from the LIA consist of
cosine (X) and sine (Y) components of the phasor which represents the phase
difference between the REF and UNK signals. These outputs are analog $-10$
to 10~V signals, which we digitize using a National Instruments PCI-6034E
board, a 4-channel data acquisition system with 16-bit resolution and a
total bandwidth of 200~kHz.  Board channels 1 and 2 capture the X and Y
signals, and the third board channel monitors the TTL trigger signal for
the mechanical shutter. The three signals are acquired at a rate of 50~kHz
and written to the hard drive of the ``metrology'' computer for later
processing.

\subsection{Fringe Data}

\begin{figure}
\epsscale{1.0}
\plotone{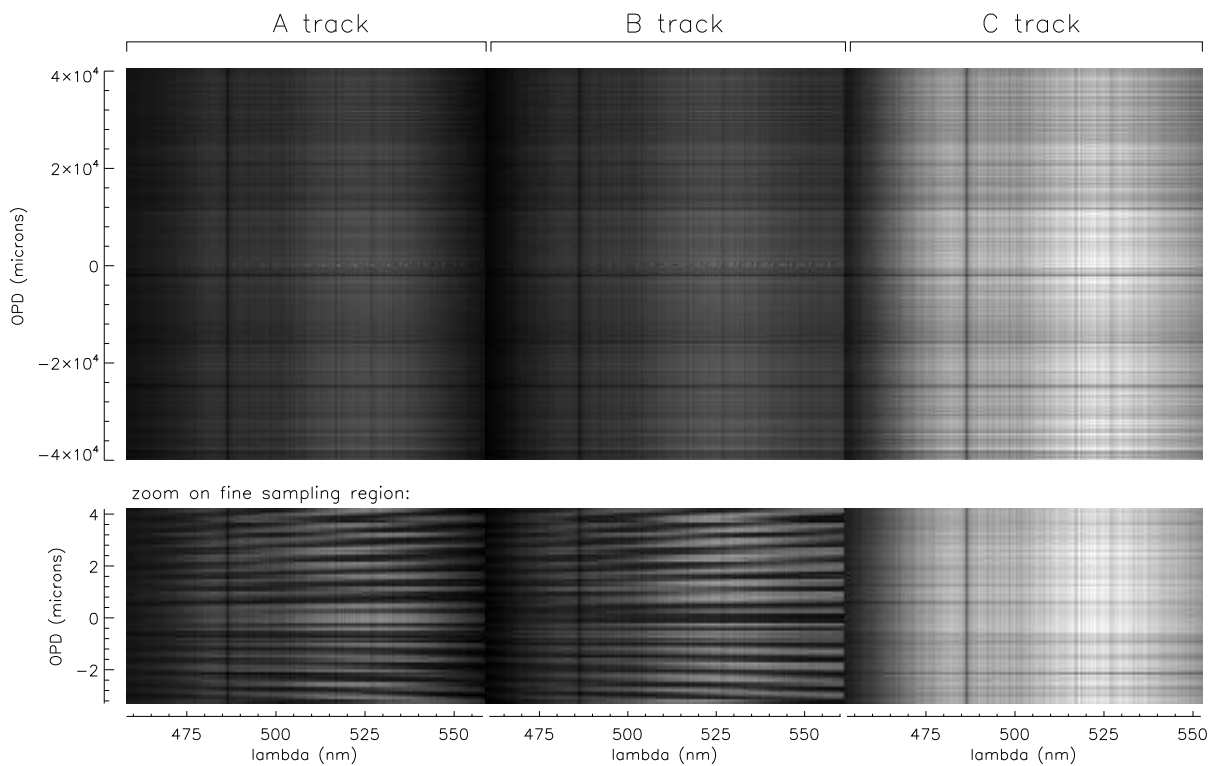}
\caption{A typical 2-d interferogram from Procyon. There are a total
of 1415 delay positions (vertical axis) and three tracks each
containing 2048 spectral channels (horizontal axis).  The top panel
shows the full delay range of $\pm 40$~mm ($60\mu$ step size), while
the bottom panel zooms in on the fine-sampling region (FSR) around the
central fringe (100~nm step size). Complementary fringing patterns are
visible in the A-track FSR and B-track FSR; the peak fringe contrast
is about 75\% of the mean level. The C track shows no fringing,
because this light does not pass through the interferometer
section. Horizontal dark bands are intensity fluctuations; these
fluctuations are removed from the A and B tracks, using the C track as
a flux reference. Vertical dark bands are stellar absorption lines,
including H$\beta$ at 486~nm, the Mg b lines around 517~nm, and the Fe
E line at 527~nm.}
\label{intf-data}
\end{figure}

The CCD and shutter are controlled by a PCI board in the ``fringe''
computer. The board activates the shutter using a TTL pulse, then
reads out three subrasters from the CCD, each of which contains one of
the dispersed spectra of the A, B, and C beams. The subrasters are
each binned vertically, so that the CCD output consists of three
``tracks,'' which we also label A, B, and C. Each track is comprised
of 2048 channels, corresponding to the columns on the CCD chip. At
each delay position during an observation (or ``scan'') of a stellar
target, we read out $3 \times 2048 = 6144$~values. Given $N_{\rm
delay}$ delay line positions, our 2-d interferogram data file contains
$6144 \times N_{\rm delay}$ total pixels. In Figure~\ref{intf-data},
we show a typical interferogram image covering a delay range of $\pm
40$~mm, centered on the central fringe, and evenly sampled every
$60\mu$ (except for the more finely-sampled region around the central
fringe). In principle, since the interferogram is symmetric around the
central fringe, we only need to sample the positive or negative half
of the delay range to derive a spectrum, but as discussed in Brault
(1985), there are significant advantages to measuring both halves of
an interferogram.  Also, since radial velocity information is not
distributed uniformly throughout delay space (see Erskine (2003) for a
detailed discussion), we might achieve better RV results by
concentrating our delay measurements on specific regions of delay
space. However, since our ultimate goal is to build a general-purpose
precision spectrometer, not just a velocimeter, we have adopted a
uniform delay sampling. We plan to explore the optimization of the
delay sampling function for specific astrophysical measurements (RV,
chemical abundances, stellar rotation, etc.) in future publications.

As described in \S 2.1, the central fringe of the fringe packet occurs
when there is zero optical path difference between the two delay
lines. At this position, all wavelengths interfere which results in
maximum signal in the A track, and minimum signal in the complementary
B track. We regularly achieve a peak fringe contrast of 75\% at this
central location. Proper reconstruction of channel spectra from the
narrowband interferograms requires that we know the location of this
central fringe location to an accuracy of a few nm. All data
acquisition scans therefore contain a ``fine sampling region'' (FSR),
within which the fringe packets are sampled every 100 nm, allowing a
sinusoid fit for each channel. The delay position at which all
channels simultaneously reach a maximum is therefore the central
fringe, and the zero point reference for the relative delay positions
measured via the metrology system.

\subsection{Computer Control Loop}

The main function of the computer control loop is to synchronize the
data streams from the fringe and metrology systems without the use of
a real-time computer operating system. The key to the synchronization
is the shutter TTL pulse, which is in the ``on'' state as long as the
shutter is open. For an interferogram consisting of $N_{\rm delay}$
delays, there are an equal number of shutter openings and closings,
which we can see in the shutter signal collected by the ``metrology''
computer DAQ card.  Combining the shutter signal and metrology LIA
phase angle data, we can solve for the optical path change from one
delay step to the next, and evaluate the fluctuation in delay (due to
vibrations, change of index of refraction of air, etc.) during each
CCD exposure. In this manner, an unambiguous path difference can be
assigned to each fringe integration, and referenced to the central
fringe location.

\subsection{Postprocessing}

The first step in postprocessing the fringe data is to obtain the
central wavelength of each spectral channel in the A and B
tracks. This is accomplished by observing an incandescent white light
source at the beginning of each observing night, which provides a
high-SNR template for the spectral bandpass of each channel. A 2-d
interferogram is collected from the white light source, using a
similar delay pattern as for stellar observations, but with finer
sampling in order to reduce ambiguity due to Fourier aliases.  The
white light narrowband interferograms are initially transformed into
spectra using a fast Lomb-Scargle periodogram algorithm (Lomb 1976;
Scargle 1982), but once the bandpass solutions are roughly known, a
more refined bandpass is calculated using the FROID algorithm
(described below).

The metrology data must then be adjusted to account for atmospheric
dispersion effects. Since the delay lines are not evacuated, light suffers
a residual optical path difference due to the change in the index of
refraction of air with wavelength, $n(\lambda)$. Furthermore, the shape of
the $n(\lambda)$ curve is a function of the air's temperature, pressure, and
relative humidity. Temperature within the instrument is kept fixed, but
pressure can vary significantly over hour timescales, so these drifts must
be factored out. Humidity is also seen to change on a seasonal basis. We
adopt Ciddor's formula (Ciddor 1996) to describe $n(\lambda, P, T, h)$.
Using this equation, the delay values for each channel can be corrected
according to:
\begin{equation}
	x_{\rm new}(\lambda)=x_{\rm old}(\lambda)
	\left[{n(\lambda)\over n(0.63281641\;{\rm \mu m})}\right]
\end{equation}
where $x_{\rm old (new)}(\lambda)$ is the optical path difference
uncorrected (corrected) for the chromatic nature of $n(\lambda)$, and
0.63281641~$\mu$m is the effective wavelength of the HeNe laser, which
sets the fringe period of the metrology data. (The base HeNe
wavelength at STP is 0.63281646~$\mu$m, but the AOMs blueshift the
laser beams by 40.0055~MHz.)  Atmospheric temperature, pressure, and
humidity corrections can yield wavelength calibration offsets
equivalent to radial velocity shifts of hundreds of meters per second,
so this correction procedure is critical for achieving the highest
possible RV accuracy.  We note that variations in the 11 kHz
modulation frequency have a negligible effect on the final velocity
calibration so long as LIA can lock on to the modulation.  Also, the
frequency fluctuation in the laser will result in small velocity
variations ($<$ 0.3 m/s) which we are currently not capable of
measuring.

In addition to the dispersion correction, the interferograms need to
be corrected for the effects of atmospheric scintillation, telescope
guiding errors, and changing atmospheric transparency. This is a
crucial step, because intensity fluctuations of the input signal might
otherwise masquerade as interferometric fringing signals (c.f. C track
in Figure 5). In theory, because the A and B interferograms are
complementary ($A+B = {\rm constant}$), we should be able to use their
sum as a normalizing factor, and distinguish flux variations from
fringes. This approach requires that the A and B spectral channels be
precisely aligned, and the system throughput as a function of
$\lambda$ for both A and B must be well-characterized. In practice, it
is easier and more reliable to use the unfringed C track to evaluate
the photometric variations, and normalize each channel of the the A
and B interferograms using the closest spectral channels in C, in case
the flux variations are not perfectly gray. 

With dispersion-adjusted and flux-corrected interferograms in hand,
and with {\it a~priori} knowledge of the central wavelength and
bandpass of each channel from the white light data, we reconstruct the
high-resolution spectrum for each channel using the Fourier
Reconstruction of Optical Interferogram Data (FROID) algorithm, which
we have developed in conjunction with the dFTS instrument. The goal of
this algorithm is to infer the narrowband channel spectrum whose
interferogram produces the best least-squares fit to the data
interferogram.  FROID has proven to be considerably more robust than
traditional Fourier algorithms when dealing with unevenly-sampled
interferogram data, and is faster as well. The mathematical
underpinnings and implementation of FROID are discussed in detail in
Appendix~B.

The individual channel spectra are then flux-weighted (using the white
light spectral bandpasses as a template) and coadded, yielding a
high-resolution, broadband spectrum for the observed target. We show
sample FROID-derived spectra for an iodine absorption cell and
assorted stars in the following sections.


\section{Performance Evaluation with Calibration Light Sources}

To demonstrate that high-resolution spectra can be reliably extracted
from the dFTS interferograms, and to evaluate the short-term and
long-term wavelength stability of the instrument, we have undertaken
observations of two calibration light sources: a molecular iodine
absorption cell and a thorium-argon emission line lamp. All
calibration measurements were taken with the dFTS {\it in situ} at the
Clay Center Observatory, not in a laboratory setting, so that we can
test the instrument performance under realistic observing
conditions. In other words, the lamp light is transported to the
guider box at the telescope and then into the dFTS through the same
light path and under the same environmental conditions as the
starlight.  The results of these experiments show that systematic
errors in the dFTS wavelength scale are on the order of a few~m/s.
This result illustrates the suitability of the instrument for
precision stellar velocimetry.

\subsection{Iodine absorption spectra}

\begin{figure}
\epsscale{1.0}
\plotone{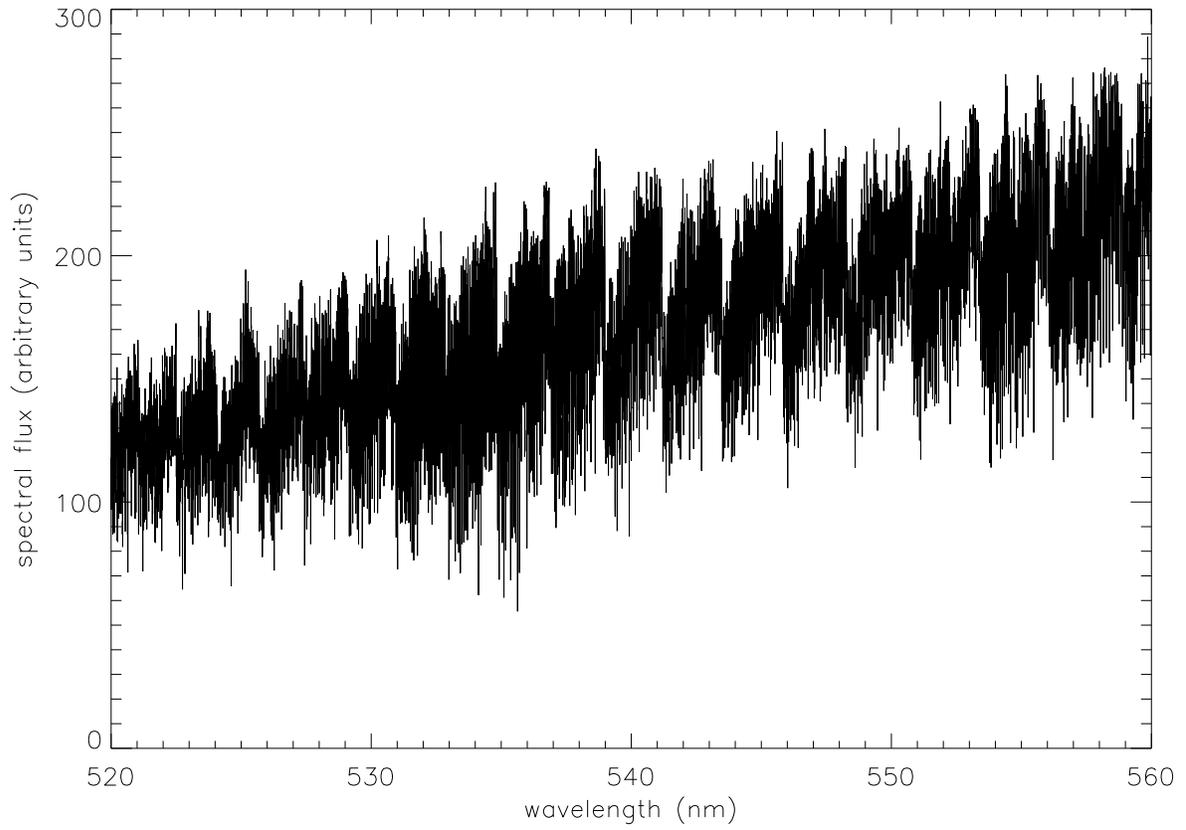}
\caption{A section of a FROID-reconstructed broadband spectrum of our
white light source passing through an iodine absorption cell. The
sawtooth pattern is due to molecular band heads. The ``noise'' is
actually many hundreds of narrow absorption lines --- see the next
figure.}
\label{iodine-spec}
\end{figure}

\begin{figure}
\epsscale{1.0}
\plotone{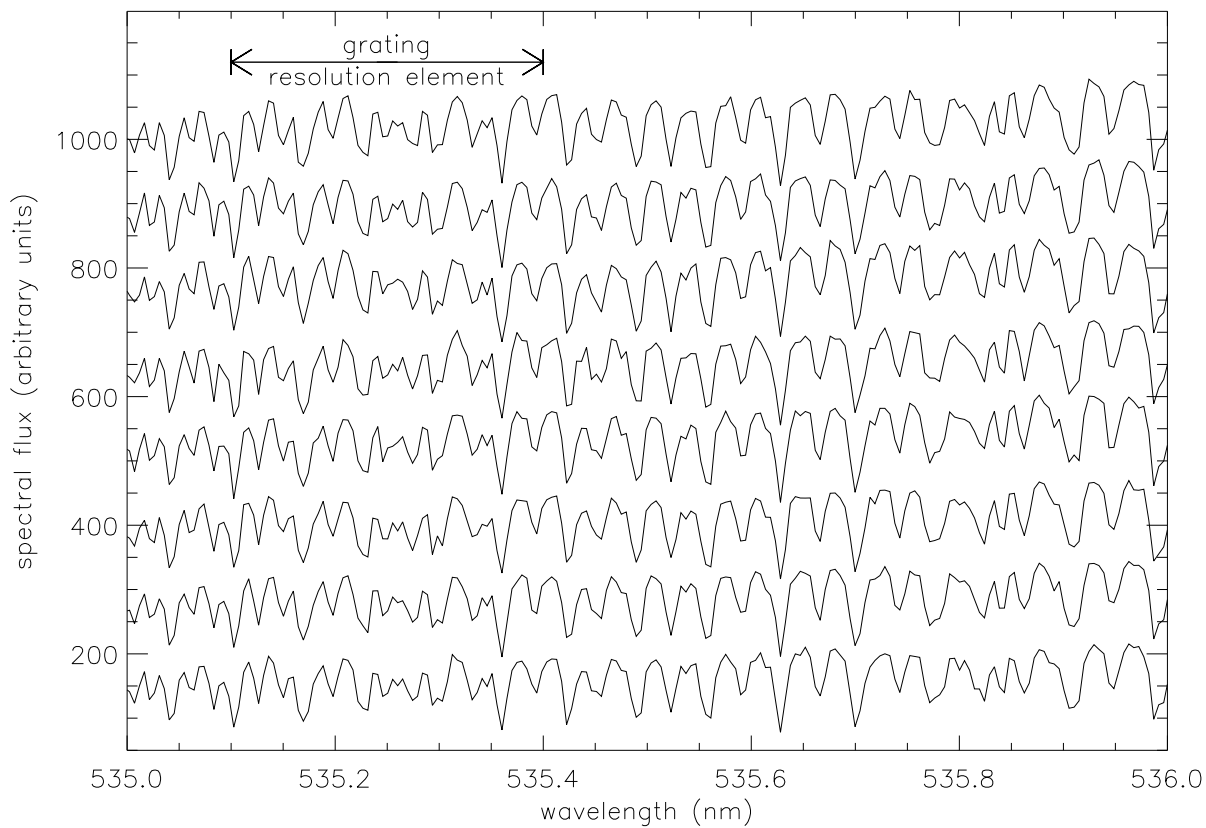}
\caption{A narrow spectral region of the previous iodine spectrum. Results from eight sequential
observations have been overplotted (with vertical offsets) to illustrate the reproducibility of the derived
spectra. The FWHM bandpass of a single dFTS channel is 0.3~nm, so all of the high-resolution content of
this spectrum, on scales of 0.1~nm and smaller, has been derived from the interference fringe patterns.}
\label{iodine-spec-zoom}
\end{figure}

As noted previously, the rich absorption spectrum of molecular iodine
serves as a wavelength reference for many of the traditional
planet-hunting programs. We use it to illustrate the spectral
resolving capabilities of the dFTS instrument and FROID algorithm. We
sent light from a 100W incandescent bulb through an iodine vapor cell
and focused it upon an optical fiber, which we coupled into the dFTS
fiber feed at the telescope.  The iodine cell was kept at a stabilized
temperature of $\approx 60\degree$C inside a small oven. We acquired
eight interferogram scans in close succession, using delay sampling
similar to that used for stellar targets. The interferograms were then
turned into broadband spectra using the FROID reduction
pipeline. Figure~\ref{iodine-spec} shows a broad region of one such
spectrum, with the molecular band heads and closely spaced absorption
lines. Figure~\ref{iodine-spec-zoom} zooms in on a smaller spectral
range, so that individual lines can be discerned, and overplots the
eight separate spectra, with vertical offset for clarity. The close
agreement in the line shapes and positions shows that the FROID
algorithm accurately reconstructs high-resolution spectra from the
sparsely-sampled interferogram data.

\subsection{Thorium-argon emission-line measurements}

For accurate velocimetry of stars at the $\sim 1$~m/s level, we must
be sure that the wavelength scale of our instrument remains constant
over extended timescales. The frequency of the HeNe laser used in our
metrology system is supposed to remain constant within $<$ 0.5~MHz
with respect to 473~THz (equivalent to an RMS scatter $<$ 0.3~m/s),
but we cannot take this specification on faith. We must also worry
about small drifts in the alignment between the metrology and
starlight beams within the interferometer, which could change the
relative path lengths and thus introduce systematic errors in our
wavelength scale.

Interferogram observations of emission-line lamps provide a straightforward
means of evaluating the stability of the dFTS. We have made time-series
observations of the rich line spectrum of a Spectral Instruments
thorium-argon discharge tube, driven by a stabilized APH 1000M Kepco power
supply. Thorium-argon spectra are often used as wavelength references in
traditional dispersive spectrographs, and therefore can be relied upon to
serve as a fixed reference source for these tests of the dFTS. We
observe our lamp source at least three times during each night that we are
collecting stellar data, as well as collecting more extensive calibration
time series on cloudy nights.

To reduce the thorium-argon scans, we select several dozen of the
strongest emission lines, extract and normalize interferograms for
each of them, and fit them with model interferograms comprising one to
four line components.  (This approach is effectively a simplified
version of the FROID algorithm, optimized for the emission-line case.)
By tracking the change in the best-fit line wavelength for each line
over a sequence of scans, we compute an RV curve for each line, and
then coadd these curves, weighted by the mean strength of each line,
to get a final RV curve. The A and B tracks of the instrument are
reduced separately as a consistency check, and to separate random and
systematic error sources; RV variation due to photon Poisson noise,
for instance, will affect A and B tracks independently, while drifts
in the reference laser wavelength or changes in alignment between the
laser and thorium-argon beams within the interferometer will show up
as correlated RV changes.

\begin{figure}
\epsscale{0.8}
\plotone{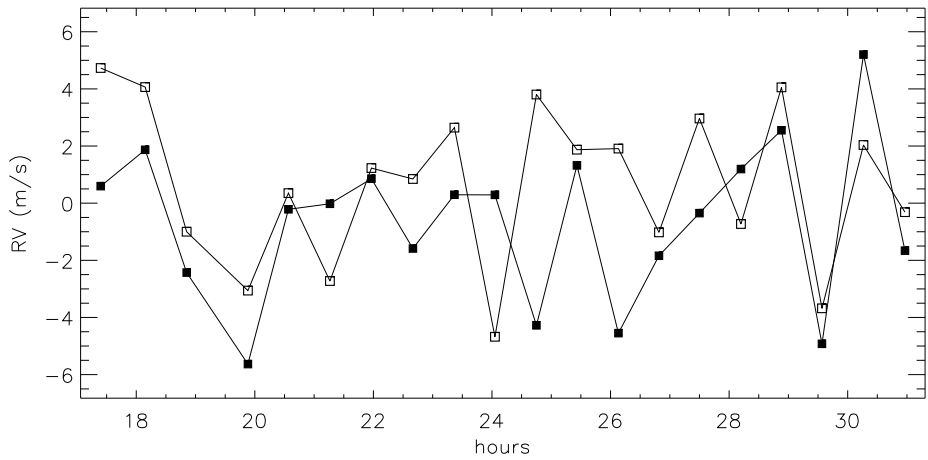}
\caption{A ``radial velocity'' curve for thorium-argon emission-line
source, for 20 separate observations during one (cloudy) night. The
instrument's A-track (solid squares) and B-track (open squares) are
reduced separately, to evaluate the relative contributions of random
photon noise and systematic error sources.}
\label{thorium-rv-1night}
\end{figure}

Figure~\ref{thorium-rv-1night} plots the RV values for 20 consecutive
thorium-argon scans spread out over 15 hours. A and B tracks are analyzed
and plotted separately. The independent time series of RV${}_{\rm A}$
and RV${}_{\rm B}$ exhibit rms scatter of 2.76~m/s and 2.74~m/s,
respectively, while the mean time series RV${}_{\rm AB} = ($RV${}_{\rm A} +
$RV${}_{\rm B})/2$ has an rms scatter of 2.29~m/s. The rms scatter of the
difference RV${}_{\rm A} - $RV${}_{\rm B}$ is 3.06~m/s, so assuming
Gaussian noise distributions, we can decompose the error signal into a
systematic component with rms 1.70~m/s and a random component with rms
2.16~m/s. Hints of the systematic error signal can be seen by eye as correlations
between the A and B RV curves in the figure.

\begin{figure}
\epsscale{0.8}
\plotone{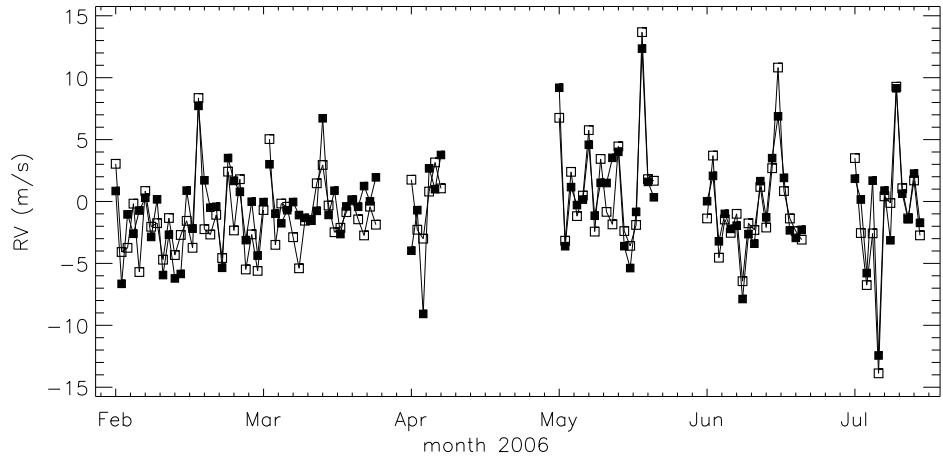}
\caption{A RV curve for thorium-argon emission-line source, spanning
six months. As in the previous figure, A-track (solid squares) and
B-track (open squares) results are plotted separately.}
\label{thorium-rv-6months}
\end{figure}

Figure~\ref{thorium-rv-6months} illustrates the RV stability of the
instrument over a longer baseline of approximately 6 months. These RV
points have been calibrated on a night-by-night basis, using
odd-numbered thorium-argon observations as a reference for the
even-numbered observations, as we would do for interleaved stellar and
thorium-argon observations. This additional calibration is necessary
to compensate for week-to-week changes in alignment between the beams
within the interferometer, which induce $\sim 10$~m/s shifts in the
uncalibrated wavelength scale. The rms scatter of the mean velocity
RV${}_{\rm AB}$ is 3.62~m/s. As before, we can decompose the separate A and
B RV data into systematic and random contributions. For the long-term data
set, we find rms(systematic)~$= 3.45$~m/s and rms(random)~$= 1.53$~m/s.

We are looking forward to improving the performance of the dFTS,
procuring a brighter lamp source, and improving the quality of the
wavelength calibration reduction software to push the wavelength
calibration to a higher level of precision.  In this way, we will be in
a better position to comment on the factors limiting the accuracy of the
velocities emerging from a dFTS.


\section{Stellar Radial Velocity Results}

\begin{figure}
\epsscale{0.6}
\plotone{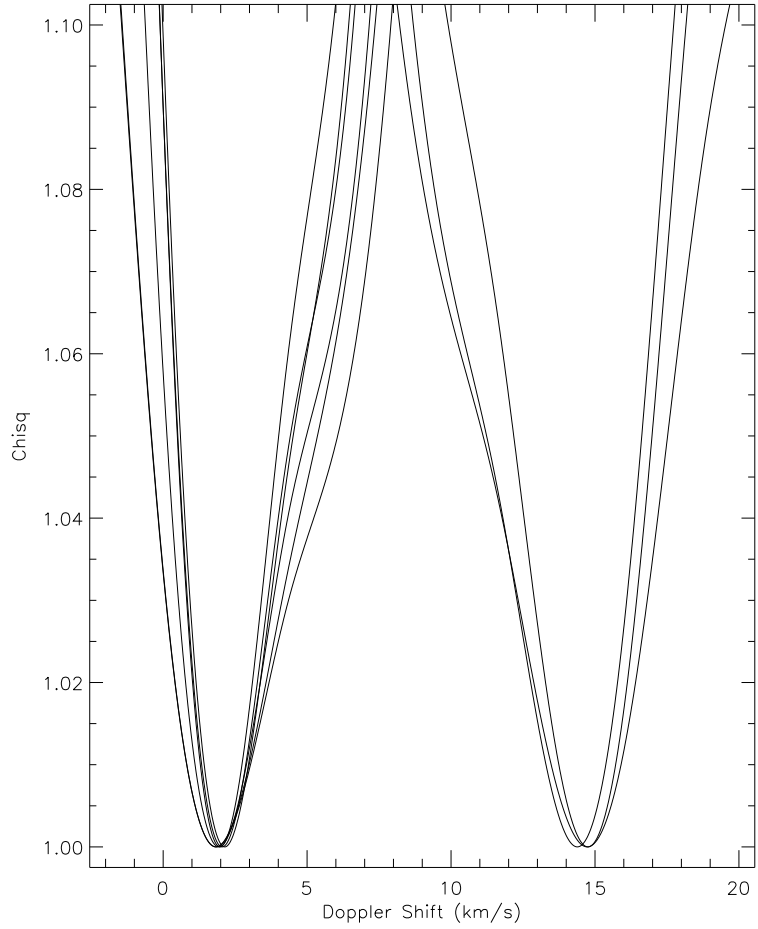}
\caption{$\chi^2$ (calculated from the cross-correlation of the
observed spectra with a template) as a function of the topocentric
radial velocity from Arcturus data for nine scans obtained on 22 May
(left group) and 20 June 2002 (right group).}
\label{arcturus-chi2}
\end{figure}

\begin{figure}
\epsscale{0.6}
\plotone{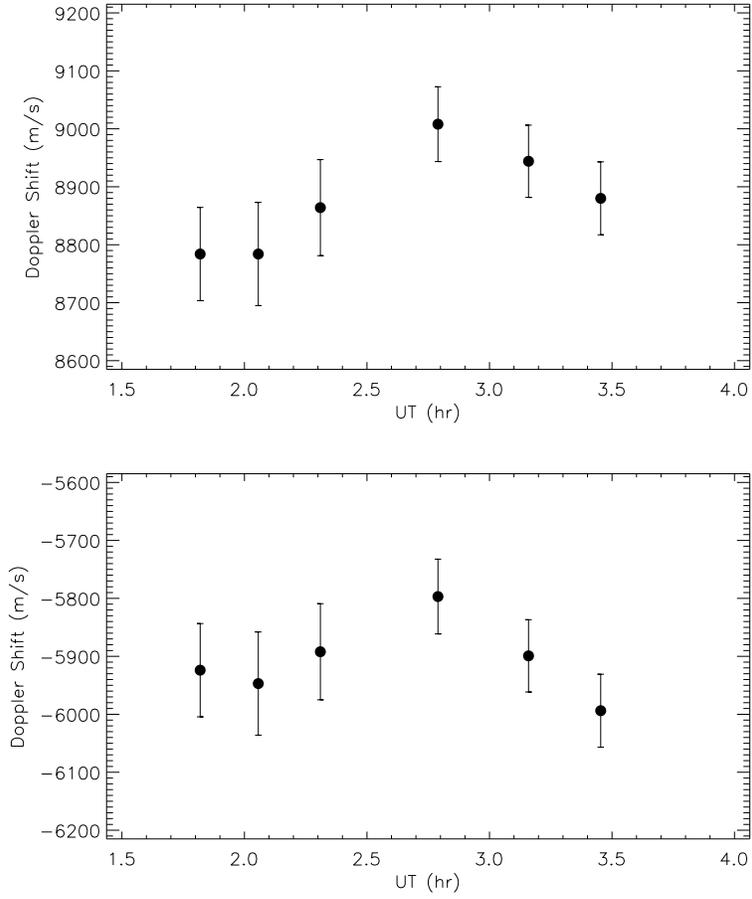}
\caption{Topocentric (top panel) and barycentric (bottom panel) radial velocity curves of Arcturus for observations in May 
of 2002. Some of the variation in topocentric RV is due to Earth's motion, but even with that removed, stellar pulsations
cause semi-periodic variations in measured barycentric RV.}
\label{arcturus-rv}
\end{figure}

Initial stellar observations were made on the grounds of the US Naval
Observatory, using a Celestron 11-inch telescope to feed a long fiber
leading inside to our optics laboratory, where the dFTS prototype was
located. We observed Arcturus ($\alpha$ Boo, KIII, $m_{\rm v} =
-0.04$) on the nights of May 22 and June 20, 2002. The acquired
interferograms show good fringe contrast, and the observed spectrum is
a good match to the model spectrum (Kurucz 1994). By cross-correlating
against this template spectrum, we derive a best-fit topocentric
radial velocity (RV) for each observation. The $\chi^2$ minima
(Figure~\ref{arcturus-chi2}) are relatively sharp, and RV values
within each night are closely grouped together. The May and June
observations give significantly different mean topocentric RV values,
because of the change in the Earth's velocity vector over one month,
and the RV trend within each night (Figure~\ref{arcturus-rv}) shows
variation due to Earth's rotation.


In October 2003, we moved the dFTS instrument to the Clay Center
Observatory at the Dexter-Southfield School in Brookline, Massachusetts.
Their 25-inch DFM telescope regularly achieves excellent image quality,
with stellar FWHMs under 1~arcsec, due to an extensive dome venting system
and vibrational isolation from the building atop which the observatory is
located. This imaging performance is thus well-suited for maximizing
throughput of the fiber feed system, which brings starlight from the
telescope down to an instrument room located underneath the dome. After a
protracted commissioning period, we started regular stellar observations in
July of 2005, with a particular focus on spectroscopic binaries and
exoplanet systems. The parameters for our primary targets are listed in
Table~\ref{star-list}, and radial velocity results are detailed below.

We used two different analysis pathways to measure stellar radial
velocities from our interferometric observations. The first pathway
uses the traditional technique of spectral cross-correlation. We
derived high-resolution spectra from our interferogram data using the
FROID algorithm, as described previously, and then performed a dual
cross-correlation, simultaneously comparing the A-track and B-track
spectra for a given stellar observation to a template spectrum
appropriate for the star's spectral type. Initial templates were drawn
from the synthetic spectrum library of Munari {\it et al.}\ (2005), and
subsequent templates were constructed by zero-shifting, co-adding, and
smoothing all the observed spectra of a star. The best-fit topocentric
velocities were shifted into the barycentric frame using the IRAF {\it
bcvcorr} task. We estimated internal error bars for each RV
measurement by measuring the width of the $\chi^2$ minimum at the
level $\chi^2 = \chi^2_{\rm min} + 1$, which corresponds to a
$1\sigma$ error interval for the measured quantity.

As an alternative to the cross-correlation approach, we also developed an
analysis algorithm which directly compares the observed interferogram data
to synthetic interferograms derived from a template spectrum. We scan
through a range of template RVs, generating a different interferogram for each
RV value, and then measuring the quality-of-fit between model
and observed interferograms via $\chi^2$, to find the RV value that gives
the best agreement. In essence, we are performing the spectral
cross-correlation without leaving Fourier space. This synthetic
interferogram fitting (SIF) algorithm returns similar RV values to the
traditional spectral cross-correlation, and provides a largely independent
verification of our FROID results.

\subsection{Procyon}

\begin{figure}
\epsscale{0.6}
\plotone{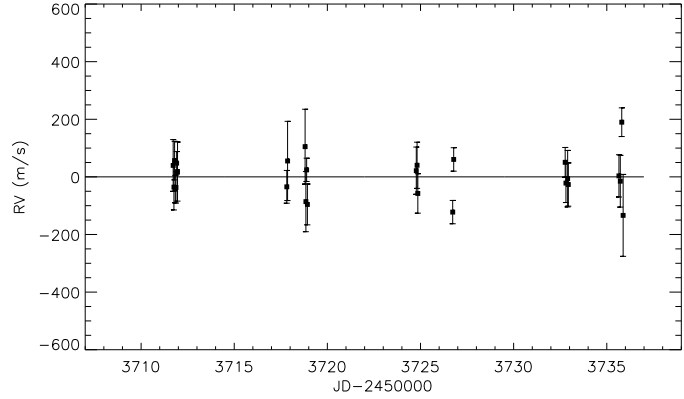}
\caption{Barycentric radial velocity curve for Procyon spanning 100 days.}
\label{procyon-rv}
\end{figure}

Procyon ($\alpha$~Canis~Minoris, HR~2943) is the brightest of our
primary stellar targets, at $m_{\rm V} \approx 0.3$. Although it is a
binary system, the orbital period is approximately 40~years, so over
short timespans, the primary serves as a velocity standard, allowing
us to check the RV stability of the instrument on a high-SNR stellar
target. Figure~\ref{procyon-rv} shows the derived barycentric radial
velocity of Procyon over a two-week interval. The RV values deviate
from their mean with an rms of 38.4~m/s. Using SNR scaling arguments,
we find that most of this RV scatter can be attributed photon
noise. To verify this conclusion, we undertook a series of
simulations, using a synthetic spectrum of a F5V star, broadening it
to match Procyon's linewidth, and then constructing a series of 16
artificial interferograms with the same delay sampling grid and mean
flux level as the actual observations. We added Poisson noise and CCD
read noise to the simulated observations, and then analyzed these
interferograms in the same fashion used for the actual data. The
resulting RV values exhibited an RMS scatter of 34.9~m/s from the
actual velocity, so it appears that photon noise is the dominant
contributor to the error budget for these observations. Future
observations with a more efficient dFTS on a larger telescope should
therefore achieve proportionately better RV accuracy.

\begin{figure}
\epsscale{0.6}
\plotone{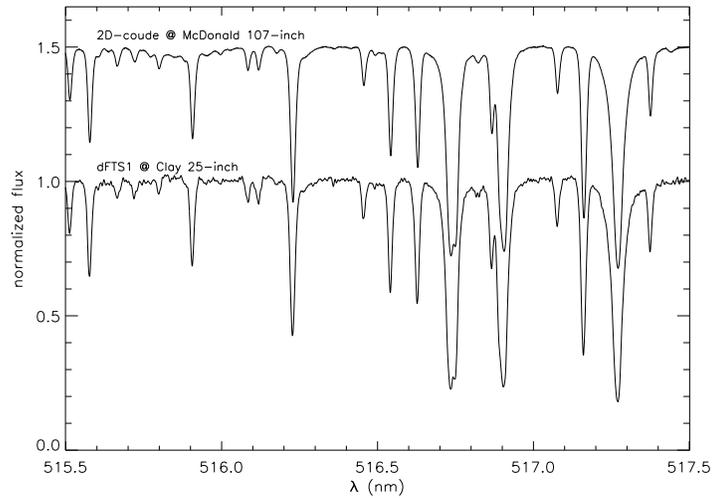}
\caption{The mean spectrum of Procyon as observed by dFTS, at a spectral resolution of $R = 50,000$, compared to the
McDonald Procyon spectral atlas of Allende Prieto et al (2002), with $R = 200,000$. The McDonald
spectrum has been offset vertically by 0.5 units for clarity.}
\label{procyon-spec}
\end{figure}

Shifting all of our observations to a common RV and coadding the
spectra, we also find that our absorption line profile shapes and
depths closely match those from the McDonald Observatory spectral
atlas of Procyon (Allende Prieto {\it et al.} 2002).
Figure~\ref{procyon-spec} illustrates this comparison over a small
subset of the full instrument bandpass.

\subsection{$\lambda$ Andromedae and $\sigma$ Geminorum}

\begin{figure}
\epsscale{0.6}
\plotone{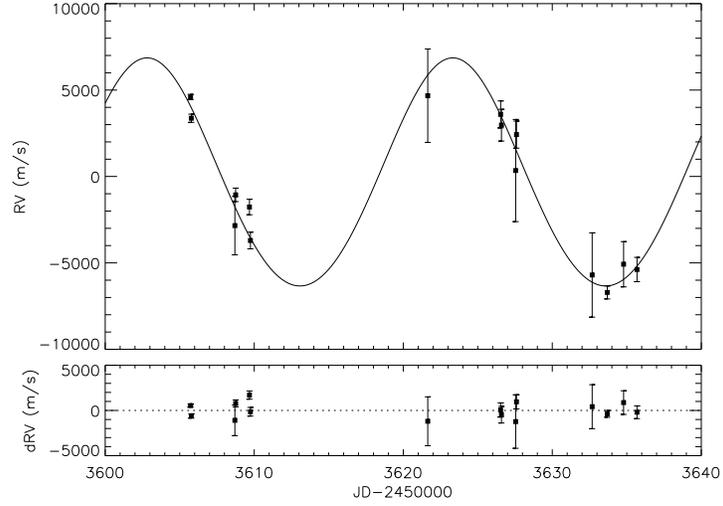}
\caption{Radial velocity curve for $\lambda$~Andromedae. The sinusoidal curve shows our best-fit orbit.}
\label{lambdaAnd-rv}
\end{figure}

To test whether the dFTS system can accurately detect RV variations in a
stellar target, we observed an assortment of spectroscopic binaries,
including $\lambda$~Andromedae (HR~8961). The RV results are plotted in
Figure~\ref{lambdaAnd-rv}. Because $\lambda$~And is nearly 30 times dimmer
than Procyon, the RV errors are correspondingly larger due to photon noise
statistics, but we still clearly detect the sinusoidal velocity variation
due to the unseen stellar companion, and our best-fit orbital solution ($P
= 20.443 \pm 0.020$~days, $K = 6557.2 \pm 35.0$~m/s) closely matches the
last published orbit (Walker 1944) with $P = 20.5212$~days and $K =
6600$~m/s. The rms scatter of our RV points from the curve is 435~m/s, a
factor of 2 larger than the mean internal error bar, perhaps indicating
that stellar variability or convective motions are contributing additional
RV noise.

\begin{figure}
\epsscale{0.6}
\plotone{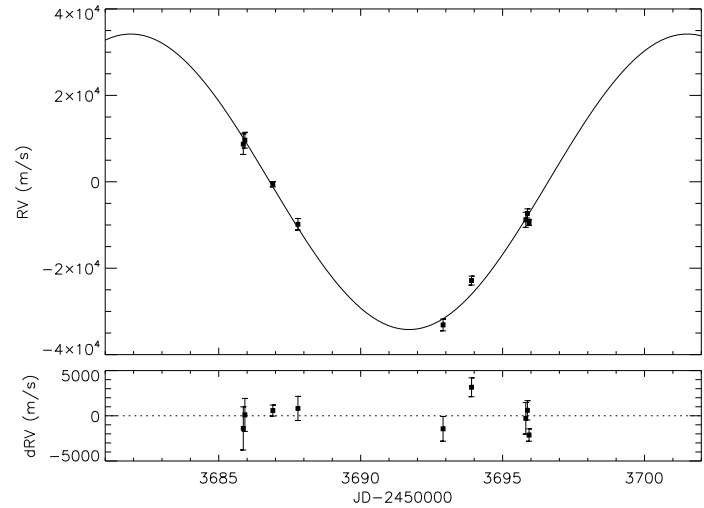}
\caption{Radial velocity curve for $\sigma$~Geminorum.}
\label{sigmaGem-rv}
\end{figure}

We also observed $\sigma$~Geminorum (HR~2973). This star rotates
somewhat faster than typical for its spectral type, perhaps due to
tidal spin-up by its unseen companion, and its photospheric absorption
lines are therefore wider, which broadens the peak of the
cross-correlation and yields greater RV uncertainty. The measured RV
points (Figure~\ref{sigmaGem-rv}) still agree well with the model RV
curve, with an rms of 463~m/s. As with $\lambda$~And, the rms scatter
is larger than the internal error estimates. We derive $P = 19.814 \pm
0.040$~days and $K = 34.3446 \pm 0.0471$~km/s, as compared to
Duemmler, Ilyin, \& Tuominen (1997), who find $P = 19.604462 \pm
0.000038$~days and $K = 34.72 \pm 0.16$~km/s. Our period measurements
are not as precise because of the limited timespan of our
observations, but the uncertainty in velocity amplitude is small,
demonstrating the precision of the dFTS.

\subsection{$\kappa$ Pegasi}

\begin{figure}
\epsscale{0.6}
\plotone{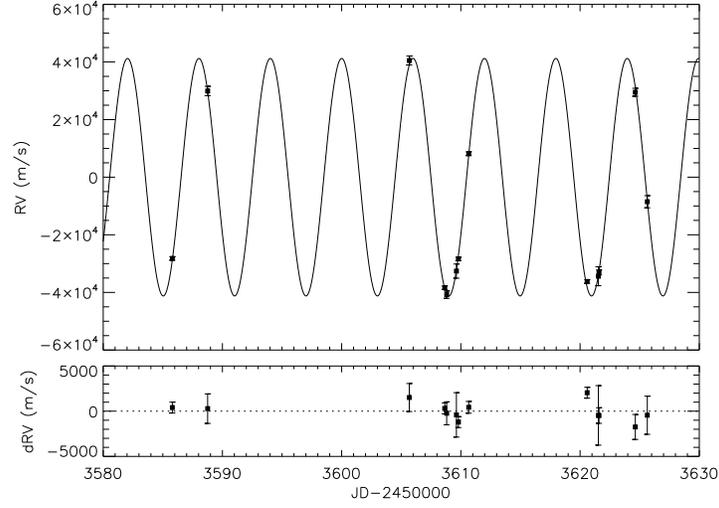}
\caption{Radial velocity curve for the short-period SB1 component of $\kappa$~Pegasi.}
\label{kappaPeg-rv}
\end{figure}

The star $\kappa$ Pegasi (HR~8315) is actually a triple system, with
two equally bright components in a 12 year orbit, and an unseen
component orbiting one of the bright stars with a 6 day period. We
observed this multiple system over a range of dates spanning several
orbital periods, and then employed a simple 2-dimensional
cross-correlation algorithm (similar to that described by Mazeh \&
Zucker 1994) to extract independent RV solutions for both bright
components.  The RV data for the sharp-lined SB1 component are
presented in Figure~\ref{kappaPeg-rv}, and once again, we find
excellent agreement between our observations and the
previously-determined ephemeris. Our points scatter around the curve
with an RMS of 990~m/s, which compares quite favorably to the 30~m/s
scatter achieved by Konacki (2005) on the same target using the Keck~I
telescope (with $\sim 250\times$ the light-gathering area of the Clay
Center telescope).

\subsection{$\tau$ Bootis}

\begin{figure}
\epsscale{0.6}
\plotone{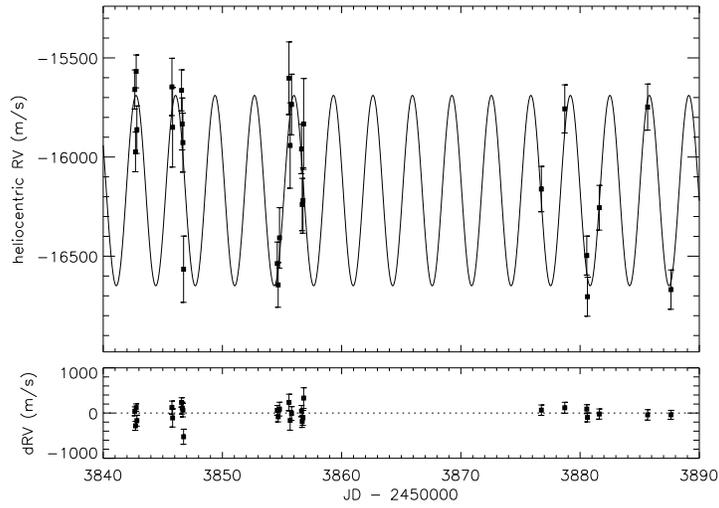}
\caption{Radial velocity curve for $\tau$~Bootis, showing the $\sim 3$ day RV oscillation
due to a massive planetary companion.}
\label{tauBoo-rv}
\end{figure}

\begin{figure}
\epsscale{0.6}
\plotone{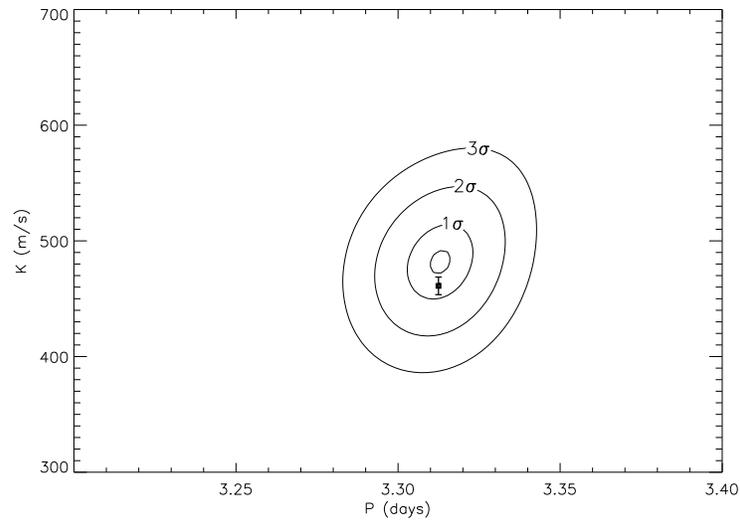}
\caption{A $\chi^2$ map of possible orbital solutions for $\tau$~Boo.
For each grid point in the $(P, K)$ parameter space, we tried to fit a
sinusoidal orbit, and evaluated the $\chi^2$ agreement between the
model and our RV data. The minimum $\chi^2$ point indicates the best
orbital solution, and the region where $\chi^2 < \chi_{\rm min}^2 + 1$
delineates the $1\sigma$ error interval for $P$ and $K$. The square
marker shows the solution ($P = 3.312463(14)$~days, $K =
461.1(7.6)$~m/s) reported by Butler {\it et al.} (2006).}
\label{tauBoo-pk}
\end{figure}

We also observed three known exoplanet host stars, in an effort to
detect the ``wobble'' in stellar RV induced by the unseen planetary
companions. We clearly detect the short-period RV variation in
$\tau$~Bootis, as shown in Figure~\ref{tauBoo-rv}, and derive $P =
3.312 \pm 0.010$~days, $K = 481.4 \pm 32.1$~m/s, nearly identical to
the orbital parameters reported in Butler {\it et al.}\ (2006). Via
$\chi^2$ fitting of sinusoidal orbits to our RV points, we construct a
map in the $(P, K)$ parameter space, showing the range of potential
orbital solutions (Figure~\ref{tauBoo-pk}). The literature solution
lies well within our $1\sigma$ error contour. This result demonstrates
that even on a small telescope, the prototype dFTS can measure stellar
RVs with sufficient accuracy to find exoplanets.

We also observed $\upsilon$~Andromedae (a 3-planet system) and
$\iota$~Draconis (a highly elliptical 1-planet system), and made
tentative planet detections in both cases. For $\upsilon$~And, our RV
data fit a weakly-constrained sinusoid with period and
velocity amplitude similar to the published parameters for planet `b,'
although there are other regions of the $(P, K)$ parameter space with
$\chi^2$ minima nearly as deep as the best-fit solution. We consider
this orbital fit to be a $1\sigma$ detection of the planet. In the
case of $\iota$~Dra, we started observations just shortly after the
periastron passage in mid-2005, so we unfortunately missed the large
anticipated ``zigzag'' in velocity.  Subsequent observations over the
following year, however, do show a monotonic change in stellar RV, as
expected from the orbital predictions of Frink {\it et al.}\ (2002).

\subsection{Performance analysis}

The photon efficiency of the dFTS instrument is low compared to
modern dispersive spectrographs. Based on count rates from stellar
observations, we estimate that the total system throughput, including
atmosphere, telescope, fiber feed, instrument optics, and CCD quantum
efficiency, is 0.7\%. Given the prototype nature of dFTS, our optics
were not optimally coated, so we take a significant hit from the $\sim
35$ optical surfaces that a photon encounters before reaching the
final focal plane.  The Canon 135 mm f/2 camera lens is particularly
bad in this regard --- we measure a throughput of 4--12\%, depending
on wavelength. Future versions of the dFTS will be able to achieve
much higher photon throughput, by utilizing antireflection-coated
lenses and high-reflectivity dielectric mirrors, minimizing the number
of fold mirrors, and replacing a photographic SLR lens with a custom
camera lens.

\begin{figure}
\epsscale{1.0}
\plotone{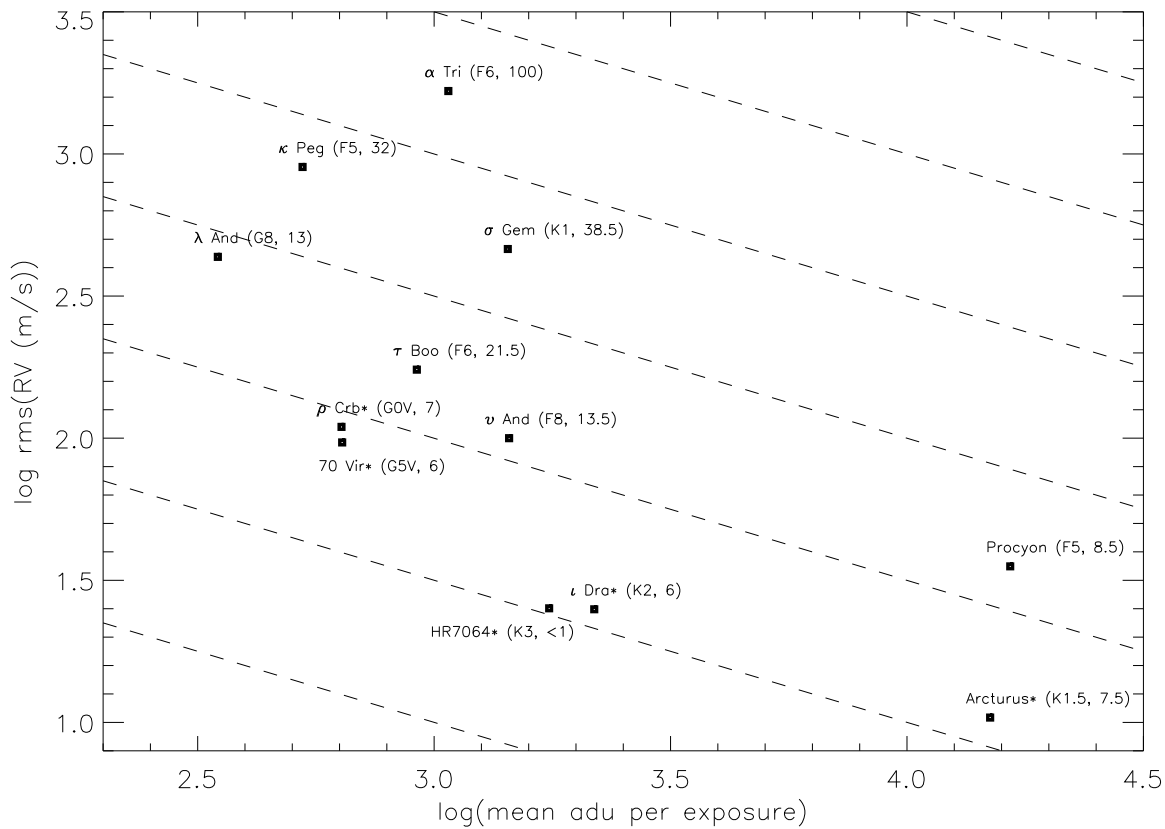}
\caption{Radial velocity precision, as estimated by rms scatter in RV
measurements, for an assortment of target stars. The horizontal axis
shows the mean number of CCD counts (adu) per channel per
interferogram exposure in the C track, and is thus proportional to
collected photons per exposure. Each star is labeled with its name,
spectral type, and typical absorption line FWHM in km/s. Asterisks
denote stars for which rms(RV) was estimated from internal error
assessments. Dashed lines show the expected trend of rms(RV)~$\propto
N_{\rm phot}^{-1/2}$.}
\label{counts-vs-rms-rv}
\end{figure}

Another metric for evaluating the performance of an instrument is its
efficiency at turning detected photons into radial velocity
information.  The RV precision on a given stellar target depends not
only on the spectrograph's performance and the photon flux, but also
on the spectral type and linewidth of the star itself. A greater
number of absorption lines, greater line depth, and sharper lines all
increase the accuracy of the RV determination.  In
Figure~\ref{counts-vs-rms-rv}, we show how measured RV precision
varies with stellar spectral type, using the rms scatter $\sigma$(RV)
from each star's best-fit orbit as a diagnostic. In addition to the
primary stellar targets discussed previously, we include several other
late-type stars which we also observed with dFTS. The $\sigma$(RV)
values for these additional stars were estimated from the internal
error bars from cross-correlation and SIF analysis, because there were
not enough RV measurements to calculate a reliable rms, or, in the
case of Arcturus, because stellar pulsations cause RV ``jitter'' well
above the measurement accuracy of a single observation. The dashed
lines in the plot indicate the expected trend of RV precision with
photon counts, assuming Poisson noise is the dominant noise,
(i.e. $\sigma(RV) \propto N_{\rm phot}^{-1/2}$.  Our stars do not lie
neatly along one of these lines. Instead, we see a strong dependence
on spectral type (which determines how many strong absorption lines
are found within our instrument's bandpass) and linewidth (which
affects the ability to accurately centroid a given spectral line).
The cooler, later-type stars with narrower linewidths yield lower
$\sigma({\rm RV})$ values, while those with broader or fewer lines are
towards the top of the plot. Among stars with similar spectral types
and linewidths, we see that the $\sigma(RV) \propto N^{-1/2}$
relationship is generally followed --- compare, for instance,
$\tau$~Boo, $\upsilon$~And, and Procyon.  By improving the photon
throughput of the instrument, and using larger-aperture telescopes, we
therefore expect to achieve stellar RV accuracies closer to the 1~m/s
level.

\section{Conclusions}

We have presented the concept, design, theory, and operation of the
dFTS instrumentation.  Our results indicate that the dFTS is a
competitive instrument for Doppler velocimetry for stellar binaries
and exoplanet detection, as well as providing high-quality,
high-resolution spectra for general stellar astrophysics.

The key to acquiring broadband optical spectra with reasonably high
sensitivity is the chromatic dispersion of the interference fringes
with a grating. This process creates a multiplexing sensitivity
gain equal to the resolving power of the grating.  By calibrating the
optical path within the interferometer with a metrology laser as the
metric, the dFTS is free to operate at wavelengths not possible with
spectrometers calibrated with iodine absorption cells.  Our use of a
blue-violet bandpass has resulted in an improved ability to convert
photons into Doppler velocities as compared to the more commonly used
red bandpass.  In addition, the spectrum from the dFTS is pure; it is
free of the absorption lines from calibration sources, it faithfully
reproduces all temporal frequencies due to symmetric and regular
sampling of the fringe packet, and it can be corrected for the
instrumental line spread function with a high degree of precision. (In
fact, one of the inspirations for the dFTS concept was the publication
of a technique for removing instrumental profiles from echelle spectra
(Butler {\it et al.}\ 1996; Valenti, Butler, \& Marcy 1995).)  At first
glance, the long term performance of the dFTS is stable at the few m/s
level.  Further investigations, which include a refined metrology
algorithm to remove the cyclic bias inherent to heterodyne metrology
systems, are required before we can comment further on the achievable
systematic error floor.

Our prototype instrument contains several conveniences and shortcuts
which are suboptimal, particularly as regards the system throughput
and photon efficiency. As a result, the stellar RV results from our
commissioning observations are limited primarily by photon noise
statistics, and performance would benefit greatly from better photon
efficiency. To this end, we are currently constructing dFTS2, which is
an improved version of the prototype dFTS described in this paper. Not
only has the photon efficiency of dFTS2 been increased by reducing
reflections and optimizing coatings, but the resolving power of the
grating has been boosted with an image slicer, resulting in further
sensitivity increases.  The optical design of dFTS2 will achieve
resolving powers of $R = 50,000$ with a 1~arcsec ``slit size'' on
telescopes as large as 3.5~meters, and yet the instrument enclosure
displaces less than 2 cubic meters, significantly smaller than an
echelle spectrograph with equivalent spectral resolving capability.
Given the advantages of combining high spectral resolution and
accuracy with small size and low cost, dFTS instruments may, with
further development, prove to be broadly useful instruments for a wide
variety of astronomical research topics.


\acknowledgements

We would like to thank many individuals who have made invaluable
contributions to the dispersed FTS project, for without their help,
this paper would not have been published. This group includes 
J.~Bangert, J.~Benson, J.~Bowles, B.~Burress, J.~Clark, T.~Corbin, 
C.~Denison, B.~Dorland, C.~Ekstrom, N.~Elias, J.~Evans, R.~Gaume,
F.~Gauss, C.~Gilbreath, L.~Ha, B.~Hicks, S.~Horner, C.~Hummel,
D.~Hutter, K.~Johnston, G.~Kaplan, T.~Klayton, R.~Millis, S.~Movit,
T.~Pauls, J.~Pohlman, T.~Rafferty, L.~Rickard, C.~Sachs, P.~Shankland,
T.~Siemers, D.~Smith, J.~Sudol, S.~Urban, N.~White, G.~Wieder, and
L.~Winter.

ARH would like to thank Landon and Lavinia Clay for their generous
support for the dFTS project over the past several years.  In
addition, ARH would like to acknowledge the excellent support offered
the dFTS project by Robert Finney and William Finney of the Dexter and
Southfield Schools. BBB is grateful for support from the NRC Research
Associateship Program and the Naval Research Laboratory, and thanks
Tom Pauls for serving as NRC mentor. We are indebted to the anonymous
referee for a thorough reading of the manuscript and many constructive
suggestions and comments. Partial support for this research was
provided by a grant from NASA in conjunction with the SIM Preparatory
Science Program (NRA 98-OSS-07).

This research has made use of the Washington Double Star Catalog
maintained at the U.S. Naval Observatory.

IRAF is distributed by the National Optical Astronomy Observatory,
which is operated by the Association of Universities for Research in
Astronomy, Inc., under cooperative agreement with the National Science
Foundation.


\appendix
\section{The sensitivity advantage of spectral multiplexing}

Consider a telescope collecting a stellar flux of $W$ photons s$^{-1}$
$\mu$m (we have expressed $W$ using units of wavenumber instead of
wavelength).  An interferogram with measurements at $N_{\rm lag}$
delays is obtained with a mean level of $W t_{\rm lag} \Delta s$
photons per lag for a given spectral channel, where $\Delta s =
s/R_{\rm g}$ is the bandwidth of the channel, $s$ is the central
wavenumber of the bandpass, $R_{\rm g}$ is the resolving power of the
grating, and $t_{\rm lag}$ is the integration time at each delay.  The
spectral resolving power implied by the maximum optical path
difference is given by:
\begin{equation}
%
R_{\rm FTS} = s~\Delta x = {s\over\delta s},
\end{equation}
where $\delta s$ is the spacing in wavenumber between adjacent
spectral intensities and $\Delta x$ is the total range of optical
delay.

In the following analysis, we consider the data from a single spectral
channel.  Since the integral of the spectral intensities over the
spectral bandwidth is equal to the intensity, $I_{\rm o}$, at
the peak of the central fringe of the interferogram, the mean spectral
intensity (i.e. the mean signal level of the spectrum) is just
$I_{\rm o}$ divided by the spectral bandwidth, $\Delta s$.  Assuming that the
fringe contrast is 100\%, then $I_{\rm o}$ is just equal to the mean
level of the interferogram, and the mean spectral intensity is:
\begin{equation}
S_{\rm S} = W t_{\rm lag},
\end{equation}
On average, the noise level in the interferogram is determined
according to Poisson statistics:
\begin{equation}
\sigma_{\rm I} = \sqrt{W t_{\rm lag} \Delta s}.
\end{equation}
Rayleigh's Theorem states that the total noise power in the spectral
and lag domains is equal:
\begin{equation}
\sigma_{\rm S} = \sigma_{\rm I} \sqrt{{\Delta x\over\Delta s}},
\end{equation}
where $\sigma_{\rm S}$ is the average spectral noise power per pixel,
and $\sigma_{\rm I}$ is the average noise power in the interferogram
per pixel.  The above expression is an approximation based on the
simplification of the integral expression, and is strictly-speaking
true only when the spectrum/interferogram is flat.  We combine
Equations (2), (3), (4) and (5) to compute the signal-to-noise ratio
in the spectrum:
\begin{equation}
SNR_{\rm S} = {S_{\rm S}\over \sigma_{\rm S}} = 
\sqrt{{W t_{\rm lag} s\over R_{\rm FTS}}}.
\end{equation}
Not surprisingly, the number of samples in the interferogram (${N_{\rm
lag}}$), is directly proportional to the number of independent
spectral values, $M$, across one channel:
\begin{equation}
%
N_{\rm lag} = {M\over\kappa} = {R_{\rm FTS}\over\kappa R_{\rm g}},
\end{equation}
and Equation (6) becomes:
\begin{equation}
SNR_{\rm S} = \sqrt{{\kappa W t_{\rm lag} N_{\rm lag} s R_{\rm g}\over
R_{\rm FTS} }}.
\end{equation}
The constant $\kappa$ is of order unity.  For the case of the
conventional FTS, $R_{\rm g}$ = 1.  Equation (8) demonstrates that
$SNR_{\rm S}$ is directly proportional to $\sqrt{R_{\rm g}}$ for a
constant integration time ($t_{\rm lag} N_{\rm lag}$), source
brightness ($W$), observing wavenumber ($s$), and spectral resolving
power ($R_{\rm FTS}$).  Sensitivity is gained with greater
multiplexing.


\section{The FROID algorithm}

We begin with two significant departures from conventional
approaches to spectral reconstruction from FTS data.  The first is
that we solve the inverse problem by taking advantage of our knowledge
about the forward problem.  Given a model spectrum,
$\tilde{I}_{m}(\{s_{j}\})$, we solve the forward problem by obtaining
the interferogram, ${I}_{m}(\{x_{i}\})$, that would result from that
model spectrum.  The inverse problem is then solved by modifying the
model spectrum until the model interferogram obtained from the forward
problem best matches the observed data, ${I}_{d}(\{x_{i}\})$.  This
approach is the opposite of the standard strategy of solving the
inverse problem by deconvolving the impulse response from
${I}_{d}(\{x_{i}\})$.  The advantages of this procedure are numerous,
including an improved ability to correct for the deleterious effects
of finite and realistic sampling, a more honest way to treat noise
statistics, and a more Bayesian treatment which will afford the ability to
incorporate prior information (Bretthorst 1988).  However, solving the
forward problem generally places higher demands on computer
processing, resulting in longer runtimes as compared to deconvolution
algorithms.

The second deviation from conventional methodology is that we choose a
continuous model spectrum rather than a spectrum that is defined only
at discrete points. Conventional methods apply Fourier Transforms to
discretely sampled data and return discretely sampled transforms. As
we show below, conventional Fourier Transforms introduce substantial
artifacts in spectra reconstructed from a sparsely sampled
interferogram.  This situation is analogous to that seen in direct
Fourier inversion of sparse aperture data from imaging spatial
interferometers, where distinguishing sidelobe structure from real
source structure is problematic.  

We present two approaches to estimating spectra from sparsely sampled
interferograms.  In the first method (hereafter, Method \# 1), we
allow only the spectral intensities to vary and assume that the
position of the central fringe is exactly known.  The second method
(hereafter, Method \# 2) also permits variation of the lag
corresponding to the central fringe in the interferogram.

\subsection{Method \# 1: Variation of the Spectral Intensities}

Use of this algorithm assumes that the location of the central fringe
is known, and that the interferogram has been shifted in delay so as
to force the zero optical path difference to occur at the exact
maximum of the central fringe. We begin with an initial model at a set
of $M$ spectral intensities, $\tilde{I}_{m}(\{s_{j}\})$, where
$\{s_j\}$ spans a wavenumber range defined by the edges of a single
narrowband channel and $M$ is the ratio of the desired resolving power
of the FTS to the resolving power of the grating.  All spectral
intensities outside this range are not free parameters, and are set to
zero.  The algorithm is robust in the sense that the final result is
very insensitive to the quality of the initial guess of the spectral intensities.

We then approximate the spectrum between $s_{j}$ and $s_{j+1}$ with a
linear interpolation between $\tilde{I}_{m}(s_{j})$ and
$\tilde{I}_{m}(s_{j+1})$.  This linear-piecewise model of the spectrum
$\tilde{I}_{m}(s)$ results in an interferogram, the value of which at
each of the $n$ lags $x_{i}$ is given by
\begin{equation}
I_{m}(x_{i}) = \sum_{j=1}^{M-1} \int_{s_{j}}^{s_{j+1}} ds~
\left[\tilde{I}_{m}(s_{j})+(s - s_{j})\Delta_{j}\right]
\cos{(2\pi x_{i}s)},
\end{equation}
where:
\begin{equation}
\Delta_{j} = \left[{\tilde{I}_{m}(s_{j+1})-
\tilde{I}_{m}(s_{j})\over s_{j+1}-s_{j}}\right].
\end{equation}
The integral can be solved analytically, so Equation (A1) becomes:
\begin{equation}
I_{m}(x_{i}) = \sum_{j=1}^{M-1}
\left[\alpha_{i,j}\tilde{I}_{m}(s_{j})+\Delta_{j}\beta_{i,j}\right],
\end{equation}
where
\begin{eqnarray}
\alpha_{i,j} & = & \left[{\sin{(2\pi x_{i}s_{j+1})}-\sin{(2\pi x_{i}s_{j})}
\over 2\pi x_{i}}\right],\\
\beta_{i,j} & = & \left[{(s_{j+1}-s_{j})\sin{(2\pi x_{i}s_{j+1})}
\over 2\pi x_{i}}\right] +
\left[{\cos{(2\pi x_{i}s_{j+1})}-\cos{(2\pi x_{i}s_{j})}\over (2\pi
x_{i})^2}\right].
\end{eqnarray}
The constants $\alpha_{i,j}$ and $\beta_{i,j}$ are defined by the
sampling in the lag and spectral spaces and need not be recalculated
for each iteration.  The mean-square difference between the model
interferogram and the data interferogram is given by:
\begin{equation}
\chi^{2} = {1\over n} \sum_{i=1}^{n}\left[I_{m}(x_{i}) -
I_{d}(x_{i})\right]^2.
\end{equation}
We now desire the model spectrum which yields a model interferogram
that best matches the data interferogram.  We can write this condition
as a set of equations minimizing $\chi^{2}$:
\begin{equation}
{\partial\chi^{2}\over\partial\tilde{I}_{m}(s_{j})} =
{2\over n} \sum_{i=1}^{n} \left[I_{m}(x_{i}) - I_{d}(x_{i})\right]
\left({\partial I_{m}(x_{i})\over\partial\tilde{I}_{m}(s_{j})}\right) = 0.
\end{equation}
To complete the statement of the problem we need to calculate the
derivatives, which are analytic:
\begin{eqnarray}
\left({\partial I_{m}(x_{i})\over\partial\tilde{I}_{m}(s_{j})}\right)  = &
\alpha_{i,1} - \left({\beta_{i,1}\over s_{2}-s_{1}}\right) & {\rm for}~j=1, \\
\left({\partial I_{m}(x_{i})\over\partial\tilde{I}_{m}(s_{j})}\right)  = &
\left({\beta_{i,j-1}\over s_{j}-s_{j-1}}\right)+
\alpha_{i,j} - \left({\beta_{i,j}\over s_{j+1}-s_{j}}\right) & {\rm for}~2
\leq j \leq M-1, \\
\left({\partial I_{m}(x_{i})\over\partial\tilde{I}_{m}(s_{j})}\right)  = &
\left({\beta_{i,M-1}\over s_{M}-s_{M-1}}\right) & {\rm for}~j=M.
\end{eqnarray}

\subsection{Method \# 2: Variation of the Spectral Intensities and Central Lag}

The interferogram is symmetric about the zero optical path difference
since the spectrum itself is real.  As a result, one would think that
determining the zero optical path difference would be trivial.
However, localizing the central lag is significantly complicated by
sparse sampling, and the inferred position is generally dependent on
the noise level, sampling of the interferogram, and finally, the
detailed shape of the spectrum.  In this paper, we assume that the
spectrum is unknown, {\it a priori}.  Additional information can be
incorporated to improve the convergence of the algorithm, depending on
the application.

In Method \# 2, we allow the central lag as well as the spectrum to
vary.  As expected, this process provides more information than Method
\# 1, at the cost of reduced numerical stability: a
noiselike initial guess for the spectrum and, more importantly, a
random guess for the central lag, are not always sufficient to
properly reconstruct the spectrum and central lag.

We proceed exactly as above, except that we explicitly write an
expression for $\chi^{2}$ as a function of $\epsilon$, the lag
corresponding to the zero optical path difference in the interferogram
[i.e., $I = I(x-\epsilon)$]:
\begin{equation}
\chi^{2} = {1\over n} \sum_{i=1}^{n}\left[I_{m}(x_{i}-\epsilon) -
I_{d}(x_{i})\right]^2.
\end{equation}
The derivatives of $\chi^{2}$ with respect to $\tilde{I}(s)$ are the
same as in Method \# 1, except that $I_{m} = I_{m}(x_{i}-\epsilon)$.
The additional derivative that is relevant to the solution of this
problem is given by:
\begin{equation}
{\partial\left(\chi^{2}\right)\over\partial\epsilon} = {2\over n}
\sum_{i=1}^{n} \left[I_{m}(x_{i}-\epsilon) - I_{d}(x_{i})\right]
\left({\partial I_{m}(x_{i}-\epsilon)\over\partial\epsilon}\right) =
0,
\end{equation}
where:
\begin{equation}
{\partial I_{m}(x_{i}-\epsilon)\over\partial\epsilon} =
{1\over x_{i}-\epsilon} \sum_{j=1}^{M-1}
\left( A_{i,j} \tilde{I}_{m}(s_{j}) + B_{i,j} \Delta_{j} \right),
\end{equation}
and:
\begin{equation}
A_{i,j} = -s_{j+1}\cos(z_{i} s_{j+1})
+s_{j}\cos(z_{i} s_{j})
+{\sin(z_{i} s_{j+1})\over z_{i}}
-{\sin(z_{i} s_{j})\over z_{i}},
%
\end{equation}
\begin{eqnarray}
\nonumber B_{i,j} & = &
s_{j}s_{j+1}\cos(z_{i} s_{j+1})
+(2 s_{j+1}-s_{j}){\sin(z_{i} s_{j+1})\over z_{i}} - s_{j} {\sin(z_{i} s_{j})
\over z_{i}}\\
& &
-s_{j+1}^{2}\cos(z_{i} s_{j+1})
+{2 \cos(z_{i} s_{j+1})\over z_{i}^{2}}
-{2 \cos(z_{i} s_{j})\over z_{i}^{2}},
\end{eqnarray}
and we have used the definition $z_{i} = 2\pi(x_{i}-\epsilon)$.  In
this implementation, $A_{i,j}$ and $B_{i,j}$ (analogous to
$\alpha_{i,j}$ and $\beta_{i,j}$ in Method \# 1) are functions of the
sampling in the lag and spectral domains, as well as $\epsilon$.



\newpage

\begin{deluxetable}{lcccccc}
\tablecaption{Primary target stars for initial dFTS RV monitoring program. \label{star-list}}
\tablewidth{0pt}
\tabletypesize{\footnotesize}
\tablehead{
				& 				&				&				&$K_1$		&spectral 		&		\\
star name			&$m_{\rm V}$ 	&binarity			&orbital period		&(km/s)		&type			&$N_{\rm obs}$ }
\startdata
Procyon			&0.3				&visual double		&40 years		&\nodata		&F5{\sc$\;$iv--v}	&25		\\
$\lambda$~And	&3.9				&SB1			&\n 20.52 days	&\n 6.6		&G8{\sc$\;$iii}	&20		\\
$\sigma$~Gem		&4.3				&SB1			&\n 19.61 days	&34.2		&K1{\sc$\;$iii}	&13		\\
$\kappa$~Peg		&4.2				&triple			&\n\n 5.97 days	&42.0		&F5{\sc$\;$iv}	&18		\\
$\tau$~Boo		&4.5				&1~planet			&\n \n 3.31 days	&\n 0.461	&F6{\sc$\;$iv}	&28		\\
\enddata
\tablecomments{Orbit parameters from Pourbaix {\it et al.}\ (2004) and Butler {\it et al.}\ (2006)}
\end{deluxetable}


\newpage

\clearpage

\clearpage

\end{document}